# Gate control of superconducting current:

# Mechanisms, parameters, and technological potential

L. Ruf[1], C. Puglia[2], T. Elalaily[3,4,5], G. De Simoni[2], F. Joint[6], M. Berke[3,4], J. Koch[1], A. Iorio[2],

S. Khorshidian[1], P. Makk[3,4], S. Gasparinetti[6], S. Csonka[3,4], W. Belzig[1],

M. Cuoco[7], F. Giazotto[2], E. Scheer[1†], A. Di Bernardo[1,8*]

1. *Department of Physics, University of Konstanz, Universitätsstraße 10, 78464 Konstanz, Germany.*
2. *NEST, Istituto Nanoscienze and Scuola Normale Superiore, 56127 Pisa, Italy.*
3. *MTA-BME Superconducting Nanoelectronics Momentum Research Group, Müegyetem rkp. 3., H-1111 Budapest, Hungary*
4. *Department of Physics, Budapest University for Technology and Economics, Müegyetem rkp. 3., H-1111 Budapest, Hungary 1111 Budapest, Hungary.*
5. *Department of Physics, Faculty of Science, Tanta University, Al-Geish St., 31527 Tanta, Gharbia, Egypt*
6. *Department of Microtechnology and Nanoscience, Chalmers University of Technology, 41296 Göteborg, Sweden.*
7. *CNR-SPIN, c/o Università degli Studi di Salerno, I-84084, via Giovanni Paolo II 132, Fisciano, Salerno, Italy.*
8. *Dipartimento di Fisica "E. R. Caianiello", Università degli Studi di Salerno, via Giovanni Paolo II 132, I-84084, Fisciano, Salerno, Italy.*

†Email: elke.scheer@uni-konstanz.de
*Email: angelo.dibernardo@uni-konstanz.de

**Abstract**
In conventional metal-oxide semiconductor (CMOS) electronics, the logic state of a device is set by a gate voltage ($V_G$). The superconducting equivalent of such effect had remained unknown until it was recently shown that a $V_G$ can tune the superconducting current (supercurrent) flowing through a nanoconstriction in a superconductor. This gate-controlled supercurrent (GCS) can lead to superconducting logics like CMOS logics, but with lower energy dissipation. The physical mechanism underlying the GCS, however, remains under debate. In this review article, we illustrate the main mechanisms proposed for the GCS, and the material and device parameters that mostly affect it based on the evidence reported. We conclude that different mechanisms are at play in the different studies reported so far. We then outline studies that can help answer open questions on the effect and achieve control over it, which is key for applications. We finally give insights into the impact that the GCS can have toward high-performance computing with low-energy dissipation and quantum technologies.

# 1. Introduction

The operation principle of modern computers based on complementary metal-oxide semiconductor (CMOS) technology relies on three-terminal transistors, which can be reversibly switched between two states via the application of a gate voltage ($V_G$) modulating the charge carrier density [1,2]. Thanks to the development of nanoscale fabrication technologies, the density of devices in CMOS circuits has steadily increased. However, the size node (currently ~ 7 nm) of transistors has recently reached a regime where further downscaling has got challenging for both technological and physical reasons [3,4]. To keep up with the constant demand for faster and more efficient electronics, alternative technologies to CMOS are therefore emerging [5]. In this context, the steadily growing power dissipation and related thermal management issues of CMOS computing platforms have become reasons of concern. Large-scale computers (supercomputers) currently under development, which can process up to $10^{18}$ floating point operations per second (flops), have in fact power requirements close to 1 GW, meaning that they need their own power plants to operate [6].

Hybrid computing architectures, which consist of low-dissipation superconducting logics combined with CMOS memories (still better than any superconducting memories), are seen as promising solution to reduce the power dissipation of supercomputers. Such hybrid architectures can reduce the power consumption of supercomputers by a factor ~ $10^2$, even after considering the cooling costs for their operation at cryogenic temperature ($T$) [6,7]. We note that cryogenic cooling of CMOS supercomputers alone would not solve the problem of their large power dissipation [6,8].

Hybrid computing systems would be easier to realize if $V_G$-controlled superconducting devices were employed for logic operations, since these can be interfaced more easily with $V_G$-controlled CMOS circuits. Nonetheless, the application of a $V_G$ to control the state of a three-terminal device made from a metallic superconductor (S) had remained unexplored for years. It is generally accepted that, unlike for doped semiconductors under a $V_G$ which enable CMOS operation, a metallic S should instead behave like any other normal metal (N), where the electric field ($E$) induced by $V_G$ is screened within the Thomas-Fermi length [9-10] (typically of a few angstroms [11]) from its surface, which would hinder any gate control in metallic systems.

Surprisingly, an experiment performed in 2018 [12] on gated superconducting Ti nanowires (Fig. 1a) has shown that these nanowires can also be switched between two different states with an applied $V_G$, similar to CMOS transistors. By measuring current versus voltage, $I(V)$, characteristics of the gated Ti nanowires, the authors of ref. [12] have shown that the superconducting critical current $I_c$ (see Box 1 for an explanation of $I_c$ and an introduction to other basic concepts of superconductivity) measured without gate voltage (i.e., at $V_G = 0$; Fig. 1b) can be reduced as $V_G$ is progressively increased, until it gets completely suppressed. Evidence for this result has been experimentally obtained by measuring a series of $I(V)$ curves at different $V_G$, as shown in Fig. 1c, and observing a progressive reduction of the vertical segment of the $I(V)$ curve for increasing $V_G$, until the $I(V)$ characteristic becomes fully linear. The presence of a vertical segment in the $I(V)$ characteristic is a signature of the device being in its superconducting state because the voltage drop $V$ measured across the device remains equal to zero despite a non-null $I$ flowing through the device (this holds true up to a maximum $I$ which defines the device $I_c$; Fig. 1b). When the $I(V)$ characteristic becomes linear upon the application of a certain $V_G$ (see Fig. 1c for $V_G = \pm 40$ V), the device has a finite resistance (i.e., it has ohmic behavior), and it gets out of the superconducting state, meaning that its $I_c$ is equal to 0.

In gated superconducting devices, the behavior just described has been also observed upon reversal of the $V_G$ polarity (i.e., when an increasingly large negative $V_G$ is applied), meaning that the $V_G$-induced suppression of $I_c$ has a bipolar nature (Fig. 1c). For this reason, to neglect the dependence of the $I_c$ suppression on the $V_G$ polarity, below we refer to $|V_G|$ without specifying the sign of the applied $V_G$.

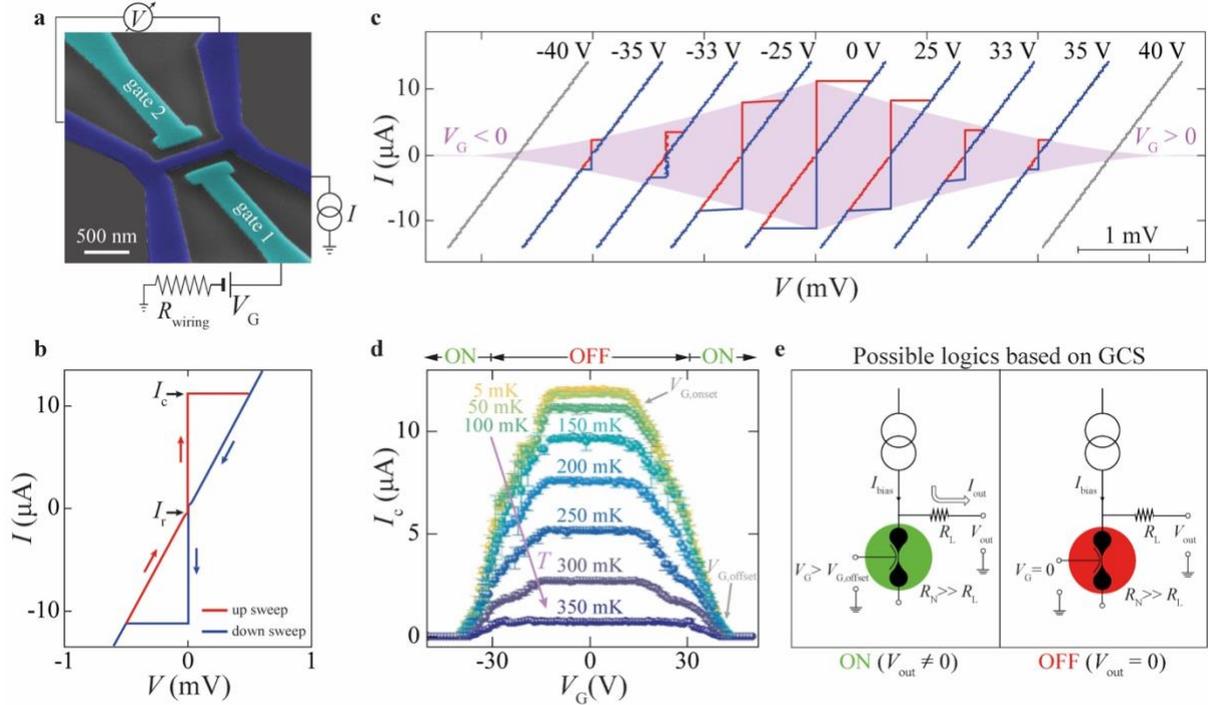

**Figure 1. Phenomenology of gate-controlled supercurrent.** (**a**-**c**) Schematic of a superconducting Ti nanowire on $SiO_2$/Si with side-gate electrodes for gate voltage $V_G$ application (a), with corresponding current versus voltage $I(V)$ characteristic measured at $V_G = 0$ (b) and at a few other positive and negative $V_G$ values (**c**), for which a progressive reduction in the critical current $I_c$ is observed as $V_G$ is increased. (**d**) $I_c$ versus $V_G$ characteristics measured at different temperatures $T$s (indicated next to each curve) showing that, independently on $T$, $I_c$ gets progressively reduced as $|V_G|$ overcomes a certain threshold $V_{G,onset}$, until it becomes fully suppressed at an even higher value ($V_{G,offset}$). (**e**) Possible circuit consisting of a gated superconducting device (represented with a colored symbol), with logic state defined by the voltage drop ($V_{out}$) on a load resistor $R_L$ connected to the superconducting device output: $V_{out} = 0$ if $V_G = 0$ (OFF state), while $V_{out} \neq 0$ if $V_G > V_{G,offset}$ (ON state). ON and OFF states correspond to the $V_G$ ranges labelled on top of horizonal axis in panel (d). Panels from (a) to (d) are adapted with permission from G. De Simoni *et al.*, Nat. Nanotechn. **13**, 802-805 (2018). Copyright 2018 Springer Nature [12].

The transition from a superconducting to a resistive state under an increasing $V_G$ in gated superconducting devices is usually not sharp. This transition can be better visualized by determining, for each $I(V)$ measured at a certain $V_G$, the corresponding $I_c$ (Fig. 1c) and then plotting the $I_c$ versus $V_G$ characteristic, as shown in Fig. 1d. The $I_c(V_G)$ curve shows that, as $V_G$ is increased, $I_c$ remains equal to its value measured at $V_G = 0$ ($I_{c0}$), until $|V_G|$ reaches a certain threshold, which we name here $V_{G,onset}$. As $|V_G|$ is further increased above $V_{G,onset}$, $I_c$ gets reduced progressively (other than abruptly) until it becomes zero at an even higher $|V_G|$ value ($V_{G,offset}$).

---

**Box 1 – Basics of superconductivity and superconducting devices**

Differently from normal conductors, where the electrical resistance gradually decreases until reaching a minimum non-zero value as temperature is reduced, superconductors (Ss) are materials where the electrical resistance vanishes, as they are cooled down below their characteristic *critical temperature* ($T_c$). In addition to zero resistance, the superconducting state is also characterized by other macroscopic properties like the Meissner-Ochsenfeld effect, which consists in the expulsion of a magnetic field from the S interior up to a certain threshold (*upper critical field*), above which superconductivity is eventually suppressed.

As a result of the vanishing resistance of a S below $T_c$, it is possible to inject a current through a S which does not dissipate energy as Joule heating. As for the magnetic field, also for the dissipationless current (*supercurrent*) injected in S can be sustained by S only up to a threshold value known as *switching supercurrent* (in this work identified with the critical current and hence denoted as $I_c$). Above this value, S turns again into a conventional dissipative conductor and its resistance becomes finite. The switching supercurrent $I_c$ decreases in amplitude as temperature is increased, since superconductivity gets weaker with increasing temperature, until vanishing at $T_c$.

If the injected bias current is progressively lowered from a value above $I_c$, meaning from a restive (dissipative) state, a S recovers its ability to support a supercurrent, although this can occur at a value of the bias current called *retrapping current* ($I_r$), which do not have to be always equal to $I_c$ but can also be lower. This hysteretic behavior stems from the Joule heating that is generated in S, after it is driven in the resistive state.

Superconductivity is a quantum phenomenon which is microscropically described, at least in most Ss made of a single chemical element, by the Bardeen-Cooper-Schrieffer (BCS) theory. The BCS theory shows that, at sufficiently low temperatures, pairs of conduction electrons can correlate in space and time forming the so-called *Cooper pairs*. The length scale over which this correlation occurs in space, which can be seen as the size of the Cooper pair, is known as superconducting coherence length $\xi$.

The formation of Cooper pairs is due to the presence of a net attractive potential, no matter how weak, stemming from the interaction of the electrons within the S material lattice. Following the BCS theory, superconductivity originates from the condensation of Cooper pairs into the same ground state. Excitations of Cooper pairs above the ground state, which are superpositions of negatively-charged electrons and positively-charged holes, are called Bogoliubov quasiparticles after Nikolay Bogoliubov. The minimum energy needed to excite a quasiparticle electron-hole couple from a Cooper pair in its groud state is the superconducting energy gap ($\Delta$). Like the supercurrent, also $\Delta$ decreases as the temperature is increased and it vanishes at $T_c$. The presence of this gap is key to enabling dissipationless transport.

Thanks to the *Josephson effect*, named after the physicist Brian Josephson who discovered it, a supercurrent can flow without any voltage applied not only in bulk Ss, but also in devices (*Josephson Junctions*, or JJs) consisting of two or more superconductors coupled by a *weak link*. Possible weak links include a thin insulating barrier (superconductor–insulator–superconductor or S-I-S JJ), a short normal metal bridge (S-N-S JJ), and a geometric constriction, with typical width of the order of few hundreds of nanometers where superconductivity is weakened (S-S'-S JJ). The latter type of weak-link is also referred to as *Dayem Bridge*.

---

A gated superconducting device is therefore in a state with zero resistance (i.e., superconducting) with $I_c \neq 0$ for $|V_G| < V_{G,offset}$ and in state with non-zero resistance (i.e., resistive) with $I_c = 0$ for $|V_G| > V_{G,\text{offset}}$. This observation has been first interpreted in ref. [12] as an effect of the $E$ field induced by the $V_G$ applied to the gate separated from the nanowire. Nonetheless, since after this first study [12], other explanations have been proposed for the GCS observation, as discussed in section 3.

To associate each of the two states ($I_c \neq 0$ and $I_c = 0$) to a different voltage output ($V_{\text{out}}$) level like in CMOS transistors, a load resistor ($R_L$) can be connected to the output of the gated superconducting device, as shown in Fig. 1e. When $|V_G| < V_{G,\text{onset}}$, a zero-voltage signal ($V_{\text{out}}$) would be measured at the $R_L$ terminals because the bias current ($I_{\text{bias}}$), which always flows through the lowest resistance path, would flow through the superconducting channel of the device connected to giving $V_{\text{out}} = 0$ (OFF state). When a $|V_G| > V_{G,\text{offset}}$ is applied, the superconducting device gets in its resistive state and, if its normal-state resistance of the device ($R_N$) is engineered to be much larger than $R_L$, then $I_{\text{bias}}$ would flow through $R_L$ giving a non-zero $V_{\text{out}}$ (ON state). $R_N$ is the resistance measured across the superconducting device in a four-point measurement setup right before the onset of its superconducting transition (or equivalently the inverse of the slope of the $I(V)$ curve when $I$ overcomes $I_c$ and $V$ becomes non-null; Fig. 1b).

Although the reversible suppression of $I_c$ under an applied $V_G$ – to which we refer hereafter as gate-controlled supercurrent (GCS) – paves the way for the development of the superconducting equivalent of CMOS logics, the physical mechanism underlying the GCS remains under debate. Understanding the physical mechanism responsible for the observation of a GCS is important from a basic science point of view. Furthermore, understanding the mechanism is also crucial to predict performance parameters such as speed, power consumption and heat generation of future logic devices based on the GCS. Similarly, studying material properties and device parameters that allow to reproduce the GCS in a systematic way are crucial steps to control this phenomenon and to develop applications based on it.

After reviewing the main features of the GCS and the different mechanisms proposed to explain it along with the evidence in support of each of them, in this article we also discuss the main material and device parameters that mostly affect the GCS, based on the studies reported to date. Throughout the review, we also outline which studies can be carried out to better understand the microscopic nature of the GCS and highlight emerging research trends in the field of the GCS that are promising for future device-oriented applications. We also discuss the open challenges that must be overcome to develop GCS-based superconducting logics with better performance than CMOS logics and other commercially-available superconducting logics like rapid single flux quantum logic. The impact of the GCS on emerging research areas like quantum computing is also briefly outlined.

## 2. Reproducibility and universal features of the GCS

Over the past 6 years since its first observation in ref. [12], several research groups have reproduced the GCS using a variety of superconductors (Ss) and device geometries. The Ss in which a GCS has been observed include elemental metallic Ss like Al [12-16], V [17,18], Ti [12,19-24], Nb [23,25-30], Ta [31], superconducting nitrides like TiN [23,32], NbN [33] and NbTiN [34], carbides like W-C [35], and non-centrosymmetric Ss like $Nb_{0.18}Re_{0.82}$ (NbRe) [36]. In only two studies [27,37], an enhancement other than a reduction in $I_c$ under $V_G$ application has been observed which, for one of these two cases [27], occurs only for a specific temperature $T$ range.

For completeness, we note that a GCS is a phenomenon commonly observed also in three-terminal S/semiconductor/S devices, where it emerges as result of a control of the charge carrier density in the semiconductor weak link via an applied $V_G$ [38-42]. Two-dimensional Ss also reveal a suppression of superconductivity under an applied $V_G$ due to a modulation in their change carrier density [43-44]. In this review article, however, we focus on gated devices made entirely of metallic Ss, which have high electron density and short Thomas-Fermi screening length. In these materials, the $V_G$-induced control of superconductivity based on the tuning of charge carrier density does not apply.

Since the metallic Ss investigated to date have different structural properties (e.g., different average grain size, degree of crystallinity etc.) and different superconducting properties ranging from higher critical temperature ($T_c$) and shorter superconducting coherence length ($\xi_S$) for Ss like NbN to lower $T_c$ and longer $\xi_S$ for Ss like Al, it may be inferred that a GCS can be observed independently on the metallic S used in a gated superconducting device. Nonetheless, even when the same S material is used, the S growth and/or the device fabrication process can determine whether a GCS is observed or not [36], as well as affect the GCS device performance (e.g., the $V_{G,offset}$ of the device), as further discussed in section 4.1.

Fig. 2a-d also shows that the GCS can be observed in devices with different geometries ranging from superconducting nanowires [12,23,32], also with a core-shell structure [14,16,31] meaning made of a S core grown onto a semiconducting shell (Fig. 2b), to Dayem bridges (see Box 1) having widths ($w_S$) of few hundreds of nanometers [15,17,19,21,25-26], to wider bridges with $w_S$ of several hundreds of nanometers [29].

The experiments performed to date show that there exist several other experimental features that are quite reproducible across devices supporting a GCS, and which can be therefore regarded as "universal signatures" of a GCS. As discussed in section 1, the GCS is mainly independent of the $V_G$ polarity, meaning that the $I_c$ suppression is approximately the same for a given $|V_G| > V_{G,offset}$ [12,14,17,20-22,26-29,35,45], as also shown by $I_c(V_G)$ characteristics at a certain $V_G$ for opposite polarities in Figs. 1 and 2. This holds true for various device realizations and gate geometries including tip-shaped [15,17,21,26] or planar [16,19,22-24] side gates, back gates [12,14], top gates [27,28], top gate with ionic liquid [25], etc. However, small asymmetries of the order of a few percent in $V_{G,offset}$ (at opposite $V_G$ polarity) have been reported by several groups [14,17,22] (see also representative asymmetric $I_c(V_G)$ curve in Supplementary Figure 1). Such small asymmetries become more pronounced when non-insulating substrates like Si are used [32], although these effects are most likely due to an asymmetric response of the substrate dependent on $V_G$ polarity (see also discussion below).

Another main feature of the GCS is its robustness against $T$ and applied magnetic field ($B$), meaning that $V_{G,offset}$ does usually not change as a function of $T$ or $B$ [12,14,17,19,21-23,26-28,31,33-36,46-47]. The $T$-independence of the GCS is evident from the $I_c(V_G)$ curves measured as a function of $T$ (for $B =$

0) that are shown in Fig. 1d and Fig. 2c. These $I_c(V_G)$ curves show that, although $I_{c0}$ gets reduced in amplitude as $T$ approaches $T_c$ due to the weakening of superconductivity, $V_{G,offset}$ remains approximately the same independently on $T$. A similar behavior is also observed when the $I_c(V_G)$ characteristics are measured as a function of $B$ (at fixed $T$), for $B$ approaching the upper critical field $B_c$ of the S (Box 1).

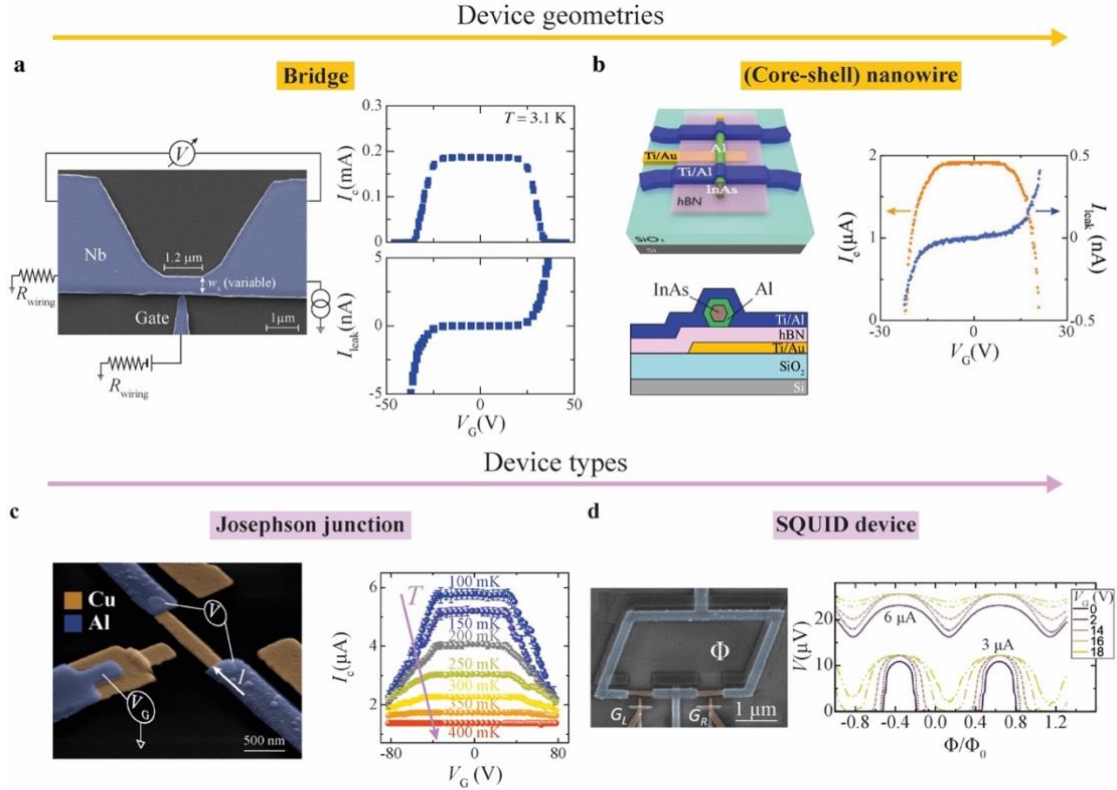

**Figure 2. Evidence for a GCS in different superconducting devices**. (**a-b**) Observation of the GCS in devices with different geometries including a bridge (a) and a nanowire (b) with device schematic shown in the left section of each panel, and critical current $I_c$ versus applied gate voltage $V_G$, $I_c(V_G)$, and leakage current $I_{leak}$ versus $V_G$, $I_{leak}(V_G)$, curves shown in the right section of each panel. Panel (a) from L. Ruf *et al.*, ACS Nano **18**, 20600 (2024) [29]; licensed under a Creative Commons Attribution (CC BY) license. Panel (b) from T. Elalaily *et al.*, Nano Lett. **21**, 9684-9690 (2021) [14]; licensed under a CC BY license. (**c**) GCS in an Al/Cu/Al Josephson junction (left panel) and corresponding $I_c(V_G)$ curves measured at different temperature $T$ (right panel). Reproduced with permission from G. De Simoni *et al.*, ACS Nano **13**, 7871-7876 (2019) [46]. Copyright 2019 American Chemical Society. (**d**) GCS devices inside an interferometer with a superconducting quantum interference device (SQUID) geometry (left panel) and corresponding voltage versus flux $\Phi$ (normalized to the flux quantum $\Phi_0$), $V(\Phi/\Phi_0)$, curve (right panel) measured under current bias exceeding the critical current $I_c$ of the SQUID of 6μA (top curves) and at current bias of 3 μA < $I_c$ (bottom curves) for different applied $V_G$ (applied to the right gate electrode of the interferometer), with $V_G$ values labelled in the panel legend. G. De Simoni *et al.*, ACS Appl. Electron. Mater. **3**, 3927-3935 (2021) [47]; licensed under a CC BY license.

The experiments performed to date also show that, upon $V_G$ application, a finite leakage ($I_{leak}$) current is always measured between the gate electrode and the electrical ground (to which one of the device terminals is also connected; Fig. 2a). Although it is often difficult to quantify the exact amount of the $I_{leak}$ that flows in the S constriction (other than along a different substrate path that does not involve the S constriction) and despite a large variation in the magnitude of $I_{leak}$ has been reported across the devices made by different groups (see discussion in section 4), $I_{leak}$ is present in devices with a side-gate or top-gate geometry where a dielectric connects the gate electrode to the S constriction. For this reason, the presence of a non-null $I_{leak}$ (independently on its magnitude) concurrent with a $V_G$ application can be considered as another typical characteristic of the GCS.

Apart from the above features, which are common to all studies reported to date, there are several other features which vary depending on the specific study considered, and which can be hence not considered as universal characteristics of the GCS.

Earlier studies, for example, suggested that a $w_S$ of the same order of magnitude as $\xi_S$ is needed for a GCS to be observed, which is the reason why side-gated devices consisting of narrow S constrictions (i.e., with $w_S$ typically up to ~ 200 nm) have been mostly studied. However, Ruf and co-workers have recently shown [29] that a GCS can also be observed in devices with $w_S \gg \xi_S$. In their study [29], the authors have also not been able to define an upper limit for $w_S$ or even to observe a progressive decrease in $V_{G,offset}$ as $w_S$ gets larger. This observation is consistent with the earlier results of ref. [12], where the authors have fabricated a series of electrodes placed at increasing distance over an elongated wire (parallel to the direction of the side gate) and found that the GCS vanishes over a length scale, which is not of the same order of magnitude as $\xi_S$ but rather comparable to the London magnetic penetration depth $\lambda_L$ (> 700 nm for Ti used as S in ref. [12]). For a S, $\lambda_L$ defines the decay length of the exponential suppression that an applied electrostatic field experiences inside S, according to London theory [48-49].

A small separation of the gate electrode from the S constriction ($d_{gate}$) seems important for a GCS to be observed. Most of the experiments performed to date have been realized with $d_{gate}$ < 100 nm [15,17,23-24,26], although the GCS has been also observed in devices with much larger $d_{gate}$ [36]. Also, it has been observed that parameters like $V_{G,offset}$ do not get reduced as $d_{gate}$ is reduced, even in the same device with two side gates placed at different $d_{gate}$ from the S constriction [31]. Therefore, although statistically one can say that a smaller $d_{gate}$ is preferable for a GCS to be observed, this cannot be considered as a strict requirement and therefore as a feature universal to all devices showing a GCS.

The last feature of the GCS proposed by earlier studies is its independence on the substrate choice [12]. In these studies, it has indeed been shown that, once a certain device geometry and S material are fixed, the GCS can be observed on different types of insulating substrates (e.g., $Al_2O_3$, $SiO_2$ etc.). Nonetheless, recent experiments [16,29] have clearly shown that the substrate can significantly affect the device performance because the applied $V_G$ can induce stress in the substrate itself, which can in turn shift the $V_{G,offset}$ of the device and its working point over time [16,29].

A list of the features typical of the GCS and discussed in this section is provided in Table 1. This list can possibly serve also as a reference for future studies on the GCS, to help confirm that the observation of a GCS is in line with previous reports.

### 2.1. Integration of GCS devices into more complex device structures

The versatility in the type of Ss and geometry of the devices supporting a GCS (Fig. 2) also suggests that GCS devices can be integrated into more complex superconducting devices and used as a knob to achieve tunability in their functionality.

| **Experimental observation** | Typical of GCS |
|---|---|
| $I_c$ suppression independent on $V_G$ polarity | YES |
| $V_{G,offset}$ mainly independent on $T$ and $B$ | YES |
| Non-null $I_{leak}$ present under applied $V_G$ | YES |
| Substrate has little effects on $V_{G,offset}$ and other device parameters | NO |
| Small device width $w_S$ (i.e., comparable to $\xi_S$) needed | NO |
| Small gate-to-channel separation $d_{gate}$ needed | YES/NO (smaller usually better) |

**Table 1.** List of experimental observations that are common to GCS devices.

To date, the GCS has been not only reproduced in S/N/S Josephson junctions (JJs), where $V_G$ is applied to the N weak link to modulate the $I_c$ of the JJ as shown in Fig. 2c [46,50], but also in more complex devices embedding a three-terminal GCS device into their layout such as superconducting resonators [13,18,30], and interferometers with a superconducting quantum interference device (SQUID) geometry [20,46].

In the case of superconducting resonators, the integration of a GCS device as a $V_G$-tunable element results in the possibility of tuning the resonant frequency $f_0$ of the resonator under $V_G$ application. This shift can be helpful to match the resonant frequency of other elements like, for example, superconducting qubits coupled to the resonator for their readout (see section 5). Nonetheless, in resonators embedding a GCS element, the shift in $f_0$ occurs alongside with a reduction in the quality factor $Q$, which corresponds to a decrease in the resonator performance. From a more technical point of view, this observation also suggests that the GCS device induces a change in the kinetic inductance of the resonator with the appearance of a real part in its impedance [13,18].

For SQUIDs, which are the devices most used for ultrasensitive magnetometry [51], the integration of a GCS device into a SQUID leads to a $V_G$-enabled tunability in the voltage versus flux, $V(\Phi)$, characteristic of the SQUID [47]. This change is different, for example, compared to that obtained by injecting a current above the $I_c$ of the SQUID. In the latter case, the SQUID operates in the fully dissipative regime and a non-null voltage is developed at its terminal independently on $\Phi$. The difference between the maximum and minimum of $V(\Phi)$, which is in turn related to the SQUID sensitivity, is fixed in this regime (Fig. 2d; top). When a $V_G$ is instead applied and the $I_c$ of the SQUID is reduced, it is possible to operate the SQUID in a regime where the maximum in $V(\Phi)$ is the same as in the resistive state, while the minimum in $V(\Phi)$ can be modulated (from zero to non-zero values) through $V_G$, meaning that the device can be still partially operated in the superconducting state (depending on $V_G$), where it has a different response to the external $\Phi$ (Fig. 2d; bottom) The integration of a GCS device in a SQUID therefore provides an additional knob, for example, when it is necessary to operate the SQUID over a wider $\Phi$ dynamic range, without reducing the overall change in voltage and in turn the SQUID sensitivity [47].

## 3. Physical mechanisms proposed for the GCS

The microscopic mechanism underlying a GCS in devices based on a metallic Ss remains under debate. The various mechanisms suggested to date to explain the GCS are those illustrated in Fig. 3, which we categorize as follows:

1) emission of high-energy electrons through vacuum relaxing into phonons and/or quasiparticles in the S, labelled in this review as "field emission" [18,23,24,28,52];

2) phonon-induced heating of the electronic system due to injection of charges into the substrate and/or into the S that can lead to an increase in the local bath temperature, in short referred to as "phonon heating" [13,14,23,32,36];

3) phase fluctuations in the S associated with an out-of-equilibrium state induced by phonons and/or high-energy electrons injected into the substrate and/or into the S, but without sizable heating of the electronic system, in short "phase fluctuations" [14,16,29,31-33,50];

4) another effect driven by the electric field associated with the applied $V_G$, henceforth called "direct field effect" [12,15,17,19-22,25-27,35-36,46-47].

We note that the borders between some of the above categories are not easy to trace. Furthermore, the assignment of a given manuscript into these four main categories is not exclusive, since the authors in some cases do not specify a single scenario active in their study, or the responsible mechanism was not yet identified at the publication date, or even several mechanisms may be at play simultaneously.

Scenario 1 explains the GCS as the result of the field emission [53] or direct tunneling of high-energy electrons from the gate electrode into the S nanoconstriction across vacuum (or vice versa for opposite $V_G$ polarity). The hot electrons injected into the S would then relax as phonons or quasiparticles inside the S, thus heating up the electronic system and hereby leading to the GCS (Fig. 3a). We note that in this scenario an $I_{leak}$ flows through the vacuum from the gate electrode into the S nanoconstriction.

The mechanisms proposed under scenarios 2 and 3 are also both triggered by $I_{leak}$. However, these are not related to vacuum tunneling, but rather to charge carriers that propagate via the substrate and thereby lose energy and excite phonons (Figs. 3b, c). The energy of the leaking charge carriers when they arrive at the S would hence be smaller on average than in scenario 1, and the phonons would be mostly created inside the substrate.

In case of scenario 2, the electronic system of the S would again be heated up because it is in contact with the warmer phonon bath, resulting in a nonequilibrium occupation of the quasiparticle system that can be described by an increased electronic temperature. According to scenario 3 instead, the superconducting condensate would be disturbed and brought into a non-equilibrium state, where the quasiparticle distribution cannot be described with an effective temperature that matches that of the bath temperature.

The difference between scenarios 1 and 2 is subtle, since scenario 1 can also include heating by phonons. However, since the energy of the charge carriers is different and the position of their decay and heat release are different, also the phenomenology related to each of these scenarios is different. Since the spatial position of charge carrier relaxation in the two scenarios is different, a way to distinguish between scenarios 1 and 2 is by the $V_G$ polarity dependence of the suppression. While in scenario 1 an asymmetric suppression is expected, scenario 2 should give rise to a symmetric suppression. Also, the distinction between scenario 2 and 3 is not easy and might just be quantitative. Experiments on devices supporting scenario 3 (see below) clearly show a GCS, but no substantial increase of the electronic $T$, despite showing similar parameters ($I_c$, $V_{G,offset}$) to those measured for devices categorized under scenario 2. Because of these subtle differences, scenarios from 1 to 3 are often referred to collectively as "leakage effects."

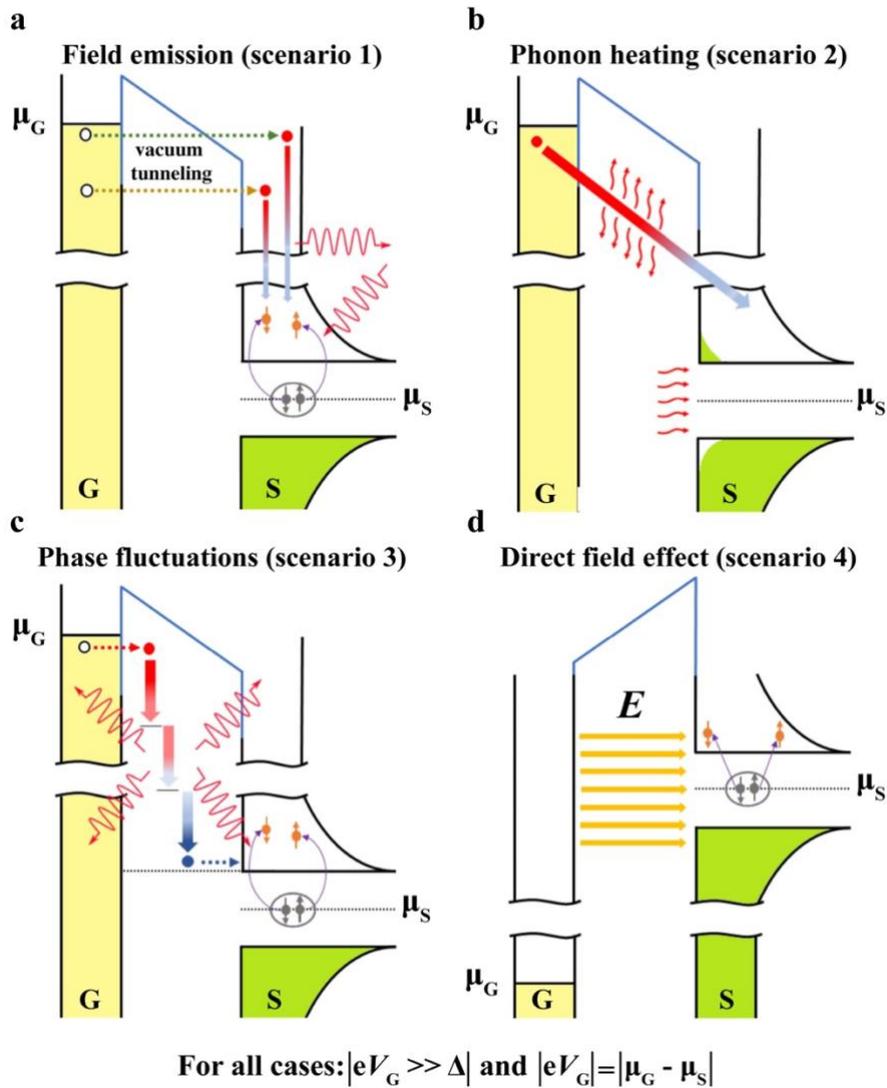

**Figure 3. Mechanisms proposed for the GCS.** Illustration of physical mechanisms proposed to explain GCS based on band diagrams of gate (G) and superconductor (S) separated by an insulator under an applied $V_G$ shifting their corresponding chemical potentials ($\mu_S$ for S and $\mu_G$ for G). The mechanisms include (**a**) tunneling across vacuum of high-energy electrons between G and S relaxing into phonons (sinusoidal arrows) and/or quasiparticles (orange dots) in S, (**b**) phonon heating (red arrows) populating quasiparticle states in S, (**c**) phase fluctuations induced by phonons triggered by high-energy electrons flowing between G and S and (**d**) $E$-field induced distortions of the superconducting phase like, for example, breaking of Cooper pairs (gray ellipses) into quasiparticles.

In contrast to the first three scenarios, scenario 4 assumes an electrostatic field effect, which can induce a GCS even in the absence of charge transfer between the S and the gate (Fig. 3d), i.e., without $I_{leak}$. Proving this scenario experimentally requires at first place ruling out scenarios 1-3, i.e., increasing the gate isolation so that the impact of $I_{lesk}$ can be excluded. Strictly speaking this task cannot be fulfilled, as long as scenarios 1-3 and the $I_{leak}$ level required for them to be at play are not clear, since there will always be a non-vanishing $I_{leak}$ associated with $V_G$, unless the resistance between the gate and the S channel increases to an unphysically infinite value.

One important message of this review is that none of these four scenarios covers the multitude of the reported phenomena reported in total. Instead, the dominating mechanism varies across experiments. In particular, we will explain that, although the majority of the experiments reported by different groups other than by single groups might fall into scenarios 2 and 3, neither scenario 1 nor scenario 4 can be fully ruled out to be at the origin of the individual experimental findings reported to date.

In Table 2 we have listed the studies on the GCS reported to date with the corresponding parameters (specified below) and grouped them according to the scenario proposed to explain the GCS observation, using differently colored boxes to identify each mechanism. The boxes overlap for those studies where more than a single mechanism can be identified or where the mechanism cannot be clearly identified. Evidence for a specific mechanism often stems from specific measurements, which have not been reproduced by other groups. For each study, in Table 2 we lists the material parameters (i.e., type of S and substrate), the main steps of the fabrication process used, and other parameters measured by characterizing the device for the GCS (i.e., $V_{G,onset}$, $V_{G,offset}$ and the corresponding $I_{leak}$ at these two $V_G$ values), $d_{gate}$ and the gate type, the ratio between the power dissipated by the gate at $V_{G,offset}$ (i.e., $P_{G,offset}$) and the power dissipated by the device when in the normal state $P_N$. The relevance of these parameters is discussed in more details in the following sections.

For the experimental device parameters, we have also adopted common criteria to extract them identically from all studies (see Fig. S1 in the Supplementary Information). A more detailed Table with and additional details on the device geometry (i.e., channel length, width etc.), other experimental parameters (e.g., $I_{c0}R_N$ product), and comments on the main findings from the authors, is also reported in the Supplementary Information.

| Proposed mechanism | S type, thickness (nm) and geometry | Substrate | Fab. process | $V_{G,onset}$ (Volts) | $V_{G,offset}$ (Volts) | $I_{leak}$ @ $V_{G,onset}$ (pA) | $I_{leak}$ @ $V_{G,offset}$ (pA) | $d_{Gate}$(nm), type | $P_{G,offset}/P_N$ | Ref. |
|---|---|---|---|---|---|---|---|---|---|---|
| **Field emission** | V(30) bridge in resonator | SiO$_2$/Si | EBL, Cl$_2$, dry etching | > 25 | n/a | ~10 | n/a | ~100 (pointy, on 2 sides) | n/a | [18] |
| | Ti (30) nanowire | SiO$_2$/Si | EBL, evap., lift off | ~27 | ~37 | ~70 | ~98 | ~70 (round, 2 on same side) | 1.84 x 10$^3$ | [24] |
| | TiN (20) nanowire | Si | EBL, HBr etching | 2.6 | 5.5 | ~0.67 | ~1.37 x 10$^2$ | > 80 (narrow flat, on 2 sides) | 2.27 x 10$^{-4}$ | [23] |
| | TiN (20) nanowire | Si | EBL, HBr etching | 1.9÷2.3 | ~3.3 | 0.5÷0.6 | 0.7÷4.4 x 10$^2$ | 80, 160 (wide flat, on 2 sides) | n/a | |

| Category | Device | Substrate | Fabrication | col5 | col6 | col7 | col8 | Gate | col10 | Ref |
|---|---|---|---|---|---|---|---|---|---|---|
| | Nb (13.5) nanowire | Si | EBL, Ar/Cl$_2$ etching | 2.4 | 4.2 | ~6.21 | 8.32 x 10$^3$ | 80 (flat, on 2 sides) | 1.5 x 10$^{-3}$ | |
| | Ti (30) nanowire | Si | EBL, evap., lift off | 1.0 | 2.6 | ~0.2 | ~26.6 | 80 (flat, on 2 sides) | 8.9 x 10$^{-2}$ | |
| | Au(3 - 5)/ Nb(3 - 5) with scaning tunneling microscope (STM) setup | Si/SiO$_2$ | EBL, etching, evap. | n/a | n/a | Varying (equal to injected tunneling current) | | n/a (STM tip-to-sample gap) | n/a | [52] |
| Phonon heating | Al (30) strip in Nb resonator | Si | EBL, evap., lift off | n/a | n/a | ~1.5 x 10$^5$ for sizeable effects | | ~80 (flat, 3 on 1 side) | n/a | [13] |
| | Al (20)/ InAs nanowire | hBN/SiO$_2$ | EBL, evap., lift off | 14.1 | 20.8 | ~74.2 | ~388 | ~ 20 (back gate) | 16.4 | [14] |
| | Au(2)/ Nb(10) (bridge) | Al$_2$O$_3$ | EBL, sputt., lift off, hBN transfer | 1.02 | 2.35 | ~4.6 x 10$^4$ | ~6.7 x 10$^5$ | Nb/hBN (6) (top gate) | 5.3 x 10$^{-3}$ | [28] |
| | NbRe (20) Dayem bridge | Al$_2$O$_3$ | EBL, Ar/Cl$_2$ etching | ~41 | 65.7 | ~ 530 | ~ 6.6 x 10$^3$ | ~291 (pointy, on 1 side) | 4.05 | [36] |
| Phase fluctuations | TiN (20) nanowire | Si | EBL, HBr etching | 3.9÷5.6 | 6.1÷7.4 | 0.4÷9.4 | 0.1÷16 x 10$^3$ | 80 (1 flat, 1 side) and 10$^3$ (2 flat, 1 side) | 2.05 x 10$^{-4}$ ÷ 3.3 x 10$^{-2}$ | [32] |
| | Ta (20) /InAs nanowire | SiO$_2$/Si | EBL, evap., lift off | 5.1 ÷ 6.3 | 11.4 ÷ 13.9 | 2 ÷25 | 1.6 ÷ 8 x 10$^2$ | 65, 115 (2 opposite sides) | 2.0 x 10$^{-2}$ ÷ 8.5 x 10$^{-2}$ | [31] |
| | Al/Cu(45)/ Al junction | SiO$_2$/Si | EBL, evap., lift off | ~ 5.6 | n/a | ~ 1.2 x 10$^{-2}$ | ~ 2.9 @10 V | ~50 (wide, T-shape, one side) | n/a | [50] |
| | Al (20)/InAs nanowire | SiO$_2$/Si | EBL, evap., lift off | 4.43 | 6.25 | 275 | 3.8 x 10$^3$ | 50, 70 (2 flat, on opposite sides) | 7.36 | [16] |
| | NbN (6) nanowire | Si | Sputt., EBL, etching | ~ 1.67 | 4.2 | n/a | 14.8 x 10$^3$ | 100 ÷ 300 (several types) | n/a | [33] |
| | Nb (27) bridge | SiO$_2$/Si | EBL, sputter., lift off | 0.85 ÷ 28.5 | 1.6 ÷ 37 | 0.07 ÷ 1.4 | 0.9 ÷ 25.3 | 50 ÷ 100 (one, on 1 side) | 5.92 x 10$^{-4}$ ÷ 1.03 | [29] |
| Direct $E$ effect | Ti (30) nanowires | SiO$_2$/Si, Al$_2$O$_3$ | EBL, evap., lift off | 16.6 ÷ 26.2 | 26 ÷ 53 | 0.004 ÷ 21 | 0.006 ÷ 41 | < 100 (flat, on 2 sides) | 4.9 x 10$^{-5}$ ÷ 3.0 x 10$^{-2}$ | [12] |
| | Al (11) nanowire | SiO$_2$/Si (doped) | EBL, evap., lift off | ~ 38 | n/a | not reported | | 300 (back gate) | n/a | [12] |
| | Al/Cu (30)/Al junction | SiO$_2$/Si | EBL, evap., lift off | 37 ÷ 43 | n/a | ~ 22 | n/a | < 100 (flat, on 1 side) | n/a | [46] |
| | Ti (30) Dayem bridge | SiO$_2$/Si | EBL, evap., lift off | ~ 17 | ~ 28 | not reported | | 80 ÷ 120 (round, on 2 sides) | n/a | [19] |
| | Ti (30) interfero-meter | SiO$_2$/Si | EBL, evap., lift off | 8.9 ÷ 40.6 | n/a | ~ 10.6 (for 1 gate) | n/a | 30 ÷ 50 (pointy, 2 same side) | n/a | [20] |
| | Nb (40) Dayem bridge | Al$_2$O$_3$ | EBL, evap., lift off | ~ 13 | ~ 44 | ~ 0.8 | > 30 | 70 (pointy, on 1 side) | 4.6 x 10$^{-2}$ | [26] |
| | V (60) Dayem bridge | SiO$_2$/Si | EBL, evap., lift off | ~ 5 | ~ 8 | not reported | | 70 (pointy, on 1 side) | n/a | [17] |
| | Ti (30) Dayem bridge | Al$_2$O$_3$ | EBL, evap., lift off | ~ 12.7 | ~ 34.3 | 2.94 | 11.3 | 80 (pointy, on 1 side) | 1.89 x 10$^{-2}$ | [21] |
| | Ti (70) nanowire (supended) | n/a | 3 EBL steps, evap. | ~ 11.0 | ~ 17.5 | 1 | 1.7 | 40 (flat, on 2 sides) | 3.67 | [22] |

| | | | | | | | | | |
|---|---|---|---|---|---|---|---|---|---|
| | Al (14) Dayem bridge | $Al_2O_3$ | EBL, evap., lift off | ~ 13.8 | ~ 22.7 | n/a | < 70 | 30 (pointy, on 1 side) | 3.8 x $10^{-1}$ | [15] |
| | Al/Cu(30)/ Al interferom. | $SiO_2$/Si | EBL, evap., lift off | 14 ÷ 46 | n/a | < 1 | n/a | 45 ÷ 60 (round, 2 same side) | n/a | [46] |
| | Nb (50) Dayem bridge | $Al_2O_3$ | EBL, sputter., lift off or sputter., EBL, etching | > 1 for lift off > 5 for etched | n/a | n/a | n/a | ~ 1 x $10^5$ (to gate electrode) Ionic liquid gating | n/a | [25] |
| | W-C (45) nanowire | $SiO_2$/Si | EBL, evap. FIB | ~ 1.5 | ~ 3 | n/a | n/a | 200 (flat, on 2 sides) | n/a | [35] |
| | NbRe (20) Dayem bridge | $Al_2O_3$ | EBL, Ar/$Cl_2$ etching | ~ 30 ÷ 41 | ~ 53.7 ÷ 62 | ~ 2 x $10^2$ ÷ 2.3 x $10^3$ | ~ 2.8 x $10^3$ ÷ 2.3 x $10^4$ | ~ 312 ÷ 321 (pointy, on 1 side) | 0.64 ÷ 7.46 | [36] |
| | 3D Nb (12) Dayem bridge | $SiO_2$/Si | EBL steps w/ sputter. and lift off | 15.3 ÷ 36.6 | 22.0 ÷ 53 | n/a | ~ $10^2$ (for gate at 130 nm) | ~ 130 ÷ 165 (top gate w/ $SiO_2$) | ~ 1.78 x $10^{-1}$ | [27] |
| Increase in $I_c$ under $V_G$ | NbN (7-10) nanowire | $SiO_2$/Si | Sputter., EBL, $Ar^+$ milling | n/a | n/a | < $10^3$ (up to 80 V) | | 300 (back gate) | $SiO_2$/Si n/a | [37] |

**Table 2 - Experimental parameters of studies on the GCS grouped based on the physical mechanism proposed**. Studies where the possible mechanism is specified by the authors are listed in a single shaded colored box, whereas studies for which the mechanism at place is not exactly specified are listed within more than one box. For parameters depending on temperature, the values reported are those measured at the base temperature of the setup (typically between 5 mK and ~ 110 mK), unless otherwise specified (see more detailed Table in the Supplementary Information).

### 3.1. Tunneling of high-energy electrons through vacuum (Field emission)

Field emission (i.e., scenario 1) has been suggested by Alegria and co-workers based on measurements of the superconducting density of states (DoS) of a S nanoconstriction (made of Ti) under an applied $V_G$ [24]. To probe the DoS of a S using a tunneling device, it is necessary to fabricate a tunnel junction on top of the S of interest, which consists of an insulating layer with a N or S electrode on top of it [54]. In the case of ref. [24], the authors have used $AlO_x$ as insulator with an Al (S) electrode on top, to fabricate the tunnel device on the Ti nanoconstriction (Fig. 4a).

The reason for choosing tunneling spectroscopy in ref. [24] to study the mechanism underlying the GCS is due to the fact that tunneling spectroscopy is a well-established technique to study how different effects including proximity effects with a different material coupled to a S (e.g., a magnetic material) or phonons – the latter can be relevant as alredy discussed in GCS devices – affect superconductivity inside the S material. All these effects in fact lead to changes in the non-linear DoS of the S, which is proportional to the differential conductance $g_{TJ} = dI_{TJ}/dV_{bias}$ [54], where $I_{TJ}$ is the current measured through the tunnel junction, while $V_{bias}$ is the bias voltage applied between the tunnel probe and the S layer.

By probing the DoS by spectroscopy with a tunneling device, in ref. [24] the authors have showed that the DoS broadens as $V_G$ is increased. (Fig. 4a). This effect is ascribed to the tunneling of quasiparticles with very high energy (~ $eV_G \gg \Delta$, $\Delta$ being the superconducting gap energy) into the S, which emit phonons that can excite further quasiparticles until the phonons escape [55]. We note that a similar broadening could also arise in scenario 2, although this was not considered at the time of the study in ref. [24].

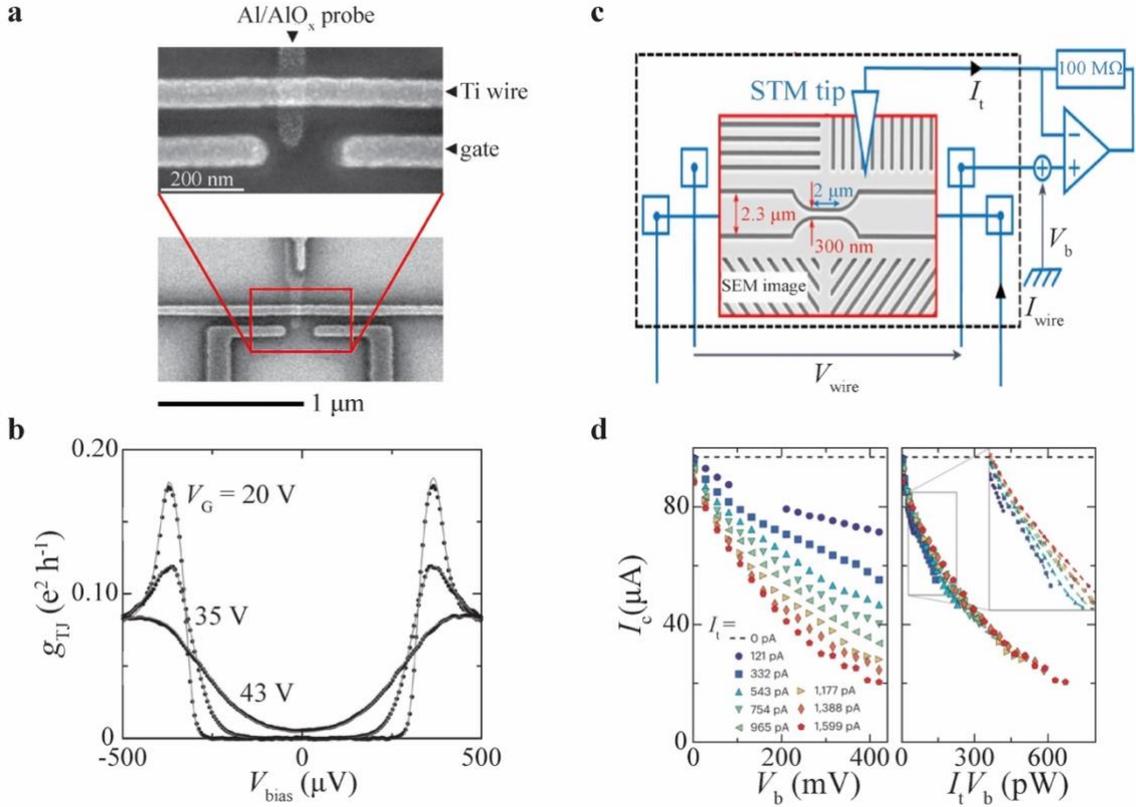

**Figure 4. Experimental evidence for tunneling of high-energy electrons.** (**a-b**) Tunneling device consisting of an Al/AlO$_x$ probe fabricated on top of a Ti nanowire with lateral gate electrodes to study the evolution of the superconducting density of states (DoS) under an applied $V_G$ (a) and corresponding density of states determined from differential conductance $g_{TJ}$ measured as a function of the bias voltage $V_{bias}$ applied between the probe and the Ti nanowire (b). Reproduced with permission from L. D. Alegria *et al.*, Nat. Nanotech. **16**, 404-408 (2021) [24]. Copyright 2021 Springer Nature. (**c-d**) Scanning tunneling microscope setup used to inject a tunnel current $I_t$ (at fixed bias voltage $V_b$) into a superconducting device (gray area) and determine its effect on the superconducting critical current $I_c$ (measured with a four-probe setup) (c) and dependence of $I_c$ on $V_b$ and on injected power $I_tV_b$ (d) for the device shown in (c). Reproduced with permission from T. Jalabert *et al.*, Nat. Phys. **19**, 956-960 (2023) [52]. Ccopyright 2023 Springer Nature.

In another experiment performed by Jalabert and co-workers [52], a scanning tunneling microscope (STM) setup has been used to study the GCS in a Nb nanowire (Fig. 4b). Although a STM can be used to measure the tunneling DoS like in ref. [24] but on a more local scale, in this study [52] the STM has not used to probe the DoS, but rather as a tool to inject quasiparticles directly from the STM tip into the underlying Nb (S) nanowire across vacuum. The motivation of the authors is that, since no solid-state tunnel junction or dielectric substrate is present in their setup, any contributions to the GCS coming from charges or photons usually excited in these materials can be excluded. As a result, the authors argue that GCS effect, which they manage to reproduce, can be only due the injection of high-energy electron from the STM tip into the Nb nanowire across vacuum. The measurements reported in ref. [52] also show that, when the energy of the quasiparticles injected e$V_G$ is larger than the S gap energy Δ, the $I_c$ suppression scales with the injected power – which is given by the product of the injected tunneling current $I_t$ and the bias voltage $V_b$ applied between the STM tip and the S nanowire (Fig. 4b). Also, the authors find that $I_c$ is almost unaffected by the injection rate of quasiparticles, which is interpreted as a signature of the quasiparticle relaxation occurring in the first tens of picoseconds after their injection. Within these tens of picoseconds, the injected quasiparticles would relax into phonons. The as-generated phonons would in turn break many Cooper pairs and generate other quasiparticles that eventually thermalize through inelastic electron-phonon and electron-electron interaction [52].

Other groups, however, give arguments against field emission as the dominating mechanism in their experiments. In some studies [14,22], for example, the authors note that field emission is inconsistent with the symmetric nature of the $I_c(V_G)$ characteristics, which is observed in most GCS devices (see section 2). According to this argument, in case field emission were responsible for the GCS, the $I_c(V_G)$ curves should be asymmetric, especially when $V_G$ is applied with a single electrode placed only on one side of the S nanoconstriction. As argued in ref. [14], this is because, while hot electrons tunneling from the gate into the S relax then in S inducing a large number of quasiparticles (and hence a significant heat load), hot electrons pulled from the S into the gate (for opposite $V_G$) heat the metal block of the gate electrode, which should in turn have a much smaller effect on S (separated by the gate through the insulator). Finite-element simulations reported in ref. [22] also show that the $I_c(V_G)$ characteristics cannot be symmetric in the case of field emission, if the device has an asymmetric geometry (i.e., the gate is only placed on one side of the nanoconstriction).

A second argument reported against field emission is based on measurements of switching current distributions (SCDs) reported in ref. [31]. For a superconducting device under $I_{bias}$, measuring the statistics of the current ($I_c$) required to switch from the device from a superconducting state (zero-voltage state) to a state of finite voltage is a very informative type of measurement. For a Josephson tunnel device, for example, the dynamics of the transition between the two states is equivalent to the process of escape of a particle from a potential well (tilted washboard potential) to a state where it runs down the potential [56-57]. At high temperature, the process is dominated by thermal activation through the barrier [58], while at low temperature it is dominated by quantum tunneling through the barrier [59]. The measurement of the SCD in these devices provides information about the escape of the phase inside the junction, and whether the dynamics of this phase escape is dominated by thermal activation or by quantum mechanical tunneling [60].

In ref. [31], like in other similar studies of the SCD in superconducting devices, the measurement of the SCD is done by biasing the device with a certain current $I$ that is ramped at a constant rate, to then measure the current value at which the device switches to a finite-voltage state. The process is repeated multiple times, to accumulate many measurements of the switching current and generate a histogram of the probability of switching at a certain $I$. According to the authors of ref. [31], field emission should in principle yield SCDs (and also average switching current $<I_c>$, if this is the only parameter measured) that are asymmetric not only when measured at the same $V_G$ with opposite polarities, but also when measured at the same power dissipated by the gate $P_G = V_G \cdot I_{leak}$ at opposite $V_G$ polarities. As explained already above, if the S nanoconstriction is grounded and $V_G$ is measured with respect to ground, for a negative $V_G$, high-energy electrons are injected from the gate into the S and hence heat the S, while, for opposite (positive) $V_G$ polarity, high-energy electrons are injected from the S into the gate and heat the gate. Under these assumptions, a stronger suppression of superconductivity should be observed, at fixed $P_G$, when the $V_G$ is negative. Nonetheless, the SCDs reported in ref. [31] are symmetric when measured at the same $P_G$ for opposite $V_G$ polarities, which is argued to be inconsistent with field emission. It is important to note that the evidence against field emission given in refs [14,22,31] does not exclude that this mechanism exists and would be the dominating one as reported in refs. [18,23-24,52].

**3.2. Heating due to phonons excited in the substrate (phonon heating)**

Scenario 2, meaning phonon-mediated heating by charge carriers leaking through the substrate, clearly triggers a GCS and a rise in the sample $T$, which can be measured concurrently with the application of $V_G$ like in ref. [13]. The main difference between reports falling under these scenarios compared to scenario 1 or 3, for example, is that the injection of an $I_{bias}$ (simulating the effect of $I_{leak}$), without an applied $V_G$, between the gate and the S nanoconstriction produces features that are identical to those that are measured just when increasing the sample $T$. For example, Catto and co-workers [13] have observed that, by recording the quality factor $Q$ and resonant frequency $f_0$ of their resonators for increasing $I_{leak}$ between the gate and an Al strip, they can reproduce the same $Q$ and $f_0$ obtained by increasing the sample $T$ (Fig. 5a). As a result, the authors conclude that the observed shift in $f_0$ and $Q$ are not due to any direct field effect (scenario 4).

In general, like in ref. [13], a good approach to understand whether scenario 2 is that mostly at play in a specific experiment, consists in tracking how a certain parameter related to the GCS (i.e., to the suppression of $I_c$) evolves for increasing $V_G$ (this in turn corresponds to an increasing $I_{leak}$), and to then compare the evolution of the same parameter with that observed by increasing the device $T$ up to the critical temperature $T_c$ of the device at $V_G = 0$. If the two trends are similar, scenario 2 is most likely the dominant mechanism toward the GCS.

Another experimental signature typically observed for devices falling under scenario 2 is a systematic shift of device parameters (> 10%) like $V_{G,onset}$ or $V_{G,offset}$ occurring as the $T$ of the sample is increased. This shift has been measured in particular for devices falling under scenario 2 which are made on non-insulating substrates like Si [23], as shown in Fig. 5d. However, the same behavior has been observed also in other devices, where the gate has poor electrical decoupling (i.e., the gate-to-channel resistance is of hundreds of kΩ or less) from the S nanoconstriction like for one the devices in ref. [36], or in devices where lab-grown $SiO_2$ has been used as dielectric to separate the gate from the S channel in a top-gate geometry [27]. If the GCS is due to $I_{leak}$-induced heating (scenario 2), it is in fact reasonable to expect that, as the sample $T$ gets closer to $T_c$ and superconductivity gets weaker, $I_c$ can be suppressed with a smaller $I_{leak}$, and hence with a lower $V_{G,offset}$.

In other devices made on commercial insulating substrates (possibly with lower density of pinholes) and with good electrical decoupling between gate and S nanoconstriction, scenario 2 has been ruled out, as explained above, by comparing the evolution of SCDs under an applied $V_G$ with that measured for increasing $T$ at $V_G = 0$. In these studies [21,31,50], it has been shown that the $V_G$ application results in much broader SCDs than those measured while increasing the sample $T$ (for the same $I_c$ suppression). For these reasons, a mechanism different from $I_{leak}$-induced heating (scenario 2) has been proposed, although phonons are still at play.

To exclude Joule heating related to $I_{leak}$ as an explanation for the GCS also in devices reported in other studies, we have also calculated the ratio between the power dissipated by the gate ($P_G$) at full suppression, $P_{G,offset} = V_{G,offset} I_{leak}$, with the power $P_N = R_N I_{c0}^2$ that the device dissipates when it switches to the resistive state, for all the studies where these parameters are available. The obtained $P_G$ values shown in Table 2 suggest that, with a few exceptions [14,22,24], $P_{G,offset}$ is usually much smaller than $P_N$, from which one can infer that the contribution from Joule heating may be minimal.

We also note that most of the devices studied to date have hysteretic current-voltage characteristics (see Fig. 1b), meaning that the transition from the superconducting to the normal state occurs at a higher absolute current (the critical current $I_c$) than the reverse transition from the normal back to the superconducting state which happens at the smaller so-called retrapping current ($I_r$). This means that, for current amplitudes between $I_r$ and $I_c$, the system is in a metastable state. It could be therefore argued that $I_r$ at $V_G = 0$ ($I_{r0}$), other than $I_{c0}$ should be considered when estimating $P_N$, which would result in a higher $P_{G,offset}/P_N$ ratio. Unfortunately, however, $I_{r0}$ values are not systematically provided. For those works where the $I_{r0}$ values have been reported, we observe a large variation in the ratio of $P_{G,offset}$ to $R_N I_{r0}^2$ across devices, independently on the scenario suggested by the authors. In particular, there are certain studies, mostly supporting scenarios 3 and 4, where $P_{G,offset}$ is smaller than $R_N I_{r0}^2$ [12,28,32], and studies where scenarios from 1 to 4 have been suggested, where $P_{G,offset}$ is either of the same order but larger [21,23,26,32] or a few order of magnitudes larger than $R_N I_{r0}^2$ [14,15,22,31]. These considerations suggest that, even in devices where relatively low $I_{leak}$ is measured at $V_{G,offset}$ (of few pA), it cannot be fully excluded that $I_{leak}$-induced Joule heating still plays a role.

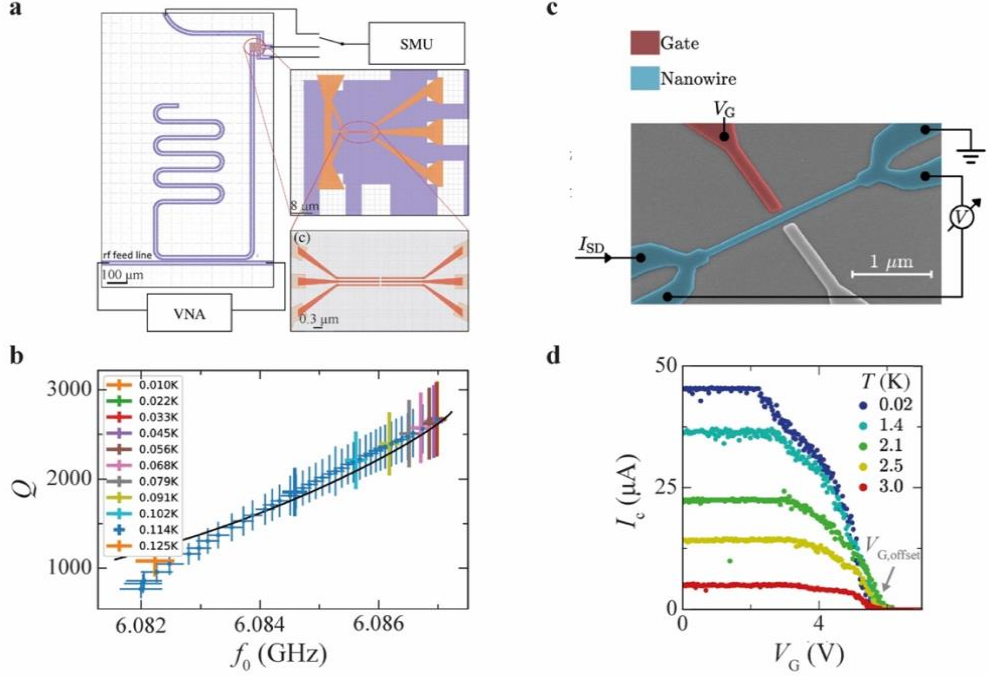

**Figure 5. Experimental evidence for phonon heating.** (**a-b**) Schematic of a co-planar waveguide resonator made of Nb (purple area) with an Al structure (orange area) used to connect the resonator to the ground plane and to study its response (a) and corresponding quality factor $Q$ versus resonant frequency $f_0$ measured at base $T$ for different values of the injected current $I_{leak}$ (blue markers) from 0 to 0.35 μA and at different temperatures as indicated by the colors in the legend (b). The data show that the dependence of $Q$ on $f_0$ for increasing $I_{leak}$ is like that measured for increasing $T$. G. Catto *et al*., Sci. Rep. **12**, 6822 (2022) [13]; licensed under a CC BY license. (**c-d**) Colored scanning electron microscope image of a gated TiN nanowire on Si substrate (c) and corresponding critical current $I_c$ versus gate voltage $V_G$ curves measured at different $T$s marked in the legend (d). Except for the $I_c(V_G)$ curve measured at $T = 20$ mK, all other curves show a progressive suppression of the $V_G$ for full $I_c$ suppression ($V_{G,offset}$) as $T$ is increased. M. F. Ritter *et al*., Nat. Commun. **12**, 1266 (2021) [23]; licensed under a CC BY license.

### 3.3. Out-of-equilibrium state due to high-energy electrons and/or phonons excited in the substrate (Phase fluctuations)

Phase fluctuations (scenario 3) are supported by other studies [29,31,50] where, even in the presence of a small $I_{leak}$ and without substantial increase of the bath $T$, the authors show that high-energy electrons in $I_{leak}$ can activate phonons in the substrate and bring the S into an out-of-equilibrium state.

In ref. [50], for example, Basset and co-authors show that even a small $I_{leak}$ of ~ 10 fA at $V_G \sim V_{G,onset}$ triggers phase fluctuations in the S constriction driving it into its resistive state. In their device, which consists of an Al/Cu/Al JJ where $V_G$ is applied to the proximitized (superconducting) Cu weak link, the authors also fabricate a tunnel probe (see Fig. 6a) to measure the DoS while applying a $V_G$ (in analogy with the device in Fig. 4a). The quasiparticle excitation spectrum probed by tunneling spectroscopy shows no traces of heating [50], and the fluctuations in $I_c$ are larger than those caused by a $T$ increase in the thermal bath [31,50] – which rules out phonon-mediated heating according to the authors.

In the same study [50], the switching dynamics of the junction is also characterized, while varying different parameter including the bath $T$, the current injected from the tunnel contact into the Cu weak link ($I_{inj}$) and the applied $V_G$. The SCDs measured at low $I_{inj}$ (corresponding to an $E$ across the tunnel barrier < 20 meV), which correspond to low-energy quasiparticles injected in the Cu weak link, are similar to those measured under increasing $T$ (see Fig. 6a). By contrast, however, the SCDs measured at a certain $V_G$ are much broader than that measured at a given $T$ or $I_{inj}$, for the same reduction in the mean switching current (i.e., mean $I_c$ value of the SCD), as shown in Fig. 6a. The analysis carried out in ref. [50] also shows that the histograms of the SCDs cannot be fitted, as $V_G$ is increased, by using an expression that only considers thermally-activated phase slips [61], but an additional term is necessary to properly fit the $V_G$-dependent SCDs. This additional term considers the effect of high-energy electrons leaking from the gate electrode, which are modelled by a Poisson distribution of temperature spikes occurring over time. The spikes are responsible for the broadening of the SCDs measured at increasing $V_G$ and induce phase fluctuations that may not necessarily switch the junction to the normal state. If the spikes overlap in time, then a global overheating can take place. The authors also argue that, in their experiment, high-energy electrons (associated with the $V_G$-induced $I_{leak}$) flow either through the substrate or via surface states.

In ref. [31] which also supports scenario 3, Elalaily et al. find that the SCDs measured at different $V_G$, are better matched (i.e., they have mean values closer to each other) when compared by the same amount of power $P_G$ dissipated by the gate (at opposite $V_G$ polarity) other than when compared by the same $V_G$ value (at opposite polarity), as shown in Fig. 6f. We note here that $P_G$ is defined as the product of $V_G$ times the $I_{leak}$ measured at the same applied $V_G$ (i.e., $P_G = V_G \cdot I_{leak}|_{V_G}$).

According to the authors of ref. [31], the dependence of the SCDs on $P_G$ and $V_G$ described above does not only show that the power dissipated by the gate $P_G$ (and hence $I_{leak}$-induced phonons) plays a crucial role toward the GCS, but it also provides evidence against a direct field effect (scenario 4), for which the SCDs measured at the same $V_G$ but with opposite polarity should be identical. Also, the SCDs measured at the same $P_G$, but with opposite $V_G$, have slightly different mean values of $I_c$ (Fig. 6f; right panel) with a dependence opposite to that expected for field emission (scenario 1). The reason behind this argument is that, for negative $V_G$, high-energy electrons would tunnel from the gate into the S nanowire, (since $V_G$ is applied between the gate and the S device which is connected to the electrical ground; see bottom-left corner of Fig. 6d). Once they land in the S nanowire, the electrons release their energy herein through relaxation. For positive $V_G$ instead, according to the device schematic in Fig. 6d, the electrons would tunnel from the S nanowire into the gate electrode, where they would also release their energy through relaxation. By comparing the two scenarios, one would except that the high-energy electrons emitted for negative $V_G$ should have a stronger impact on the suppression of superconductivity in the S nanowire, meaning that the SCD measured at negative $V_G$ (for a fixed $P_G$ dissipated by the gate) should have a lower mean value compared to the SCD measured at the same $P_G$ but for negative $V_G$. This is, however, exactly the opposite to what the authors of ref. [31] have found, as shown by the data in Fig. 6f. For this reason, scenario 1 is excluded as possible mechanism behind the GCS in the devices reported in this study [31].

In the same study [31], the authors also report that $V_{G,onset}$ and $V_{G,offset}$ change between cooldowns and that their values do not scale with the gate-to-channel distance. This is shown in Fig. 6e, where the $I_c(V_G)$ curve measured for the gate closer to the S nanowire (gate 1; Fig. 6d) shows a larger $V_{G,offset}$ than the gate placed further away from the same nanowire (gate 2; Fig. 6d). The fact that $V_{G,offset}$ does not scale with $d_{gate}$ is interpreted as additional evidence against scenario 4 [31], since a gate electrode closer to the same S constriction and made on the same dielectric substrate, should give a larger $E$ at a given $V_G$ ($E \propto V_G/d_{gate}$).

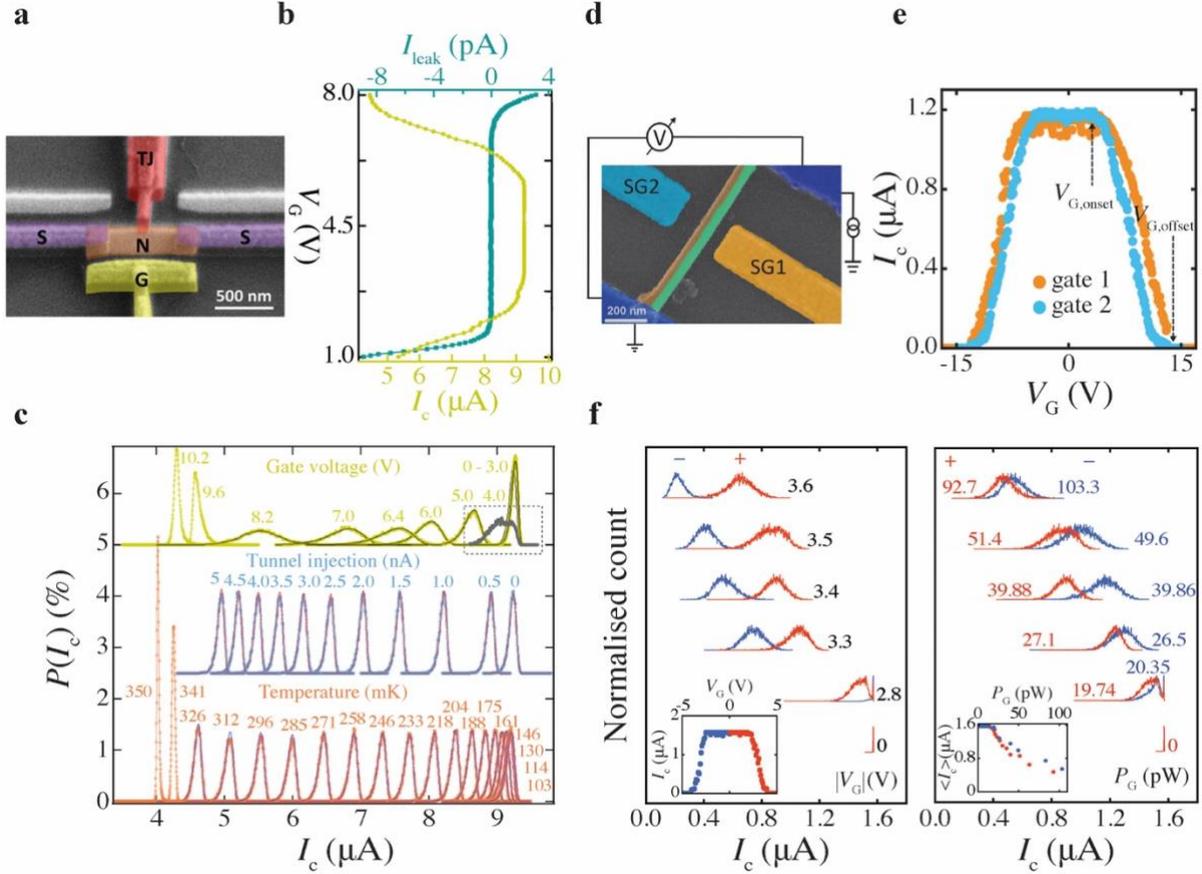

**Figure 6. Evidence for phase fluctuations.** (**a**) Colored scanning electron microscope (SEM) image of a Josephson junction consisting of two Al (S) electrodes separated by a Cu (N) weak link, with side gate (G) electrode and tunnel junction (TJ) probe of Al/AlO$_x$. (**b**) Critical current (bottom axis) $I_c$ and leakage current $I_{leak}$ (top axis) versus applied gate voltage ($V_G$) for the device in (a). (**c**) Switching current distributions (SCDs) for the device in (a) measured at different $V_G$ (yellow curves), tunnel injection current (blue curves) and temperature (red curves). The fits to the SCDs are plotted with solid lines, while raw data with symbols. Panels from (a) to (c) are adapted from J. Basset *et al.*, Phys. Rev. Res. **3**, 043169 (2021) [50]; licensed under a CC BY license. (**d-e**) Colored SEM image of core-shell superconducting nanowire made of Ta shell (on InAs core) with two gates (d), and corresponding $I_c(V_G)$ curves measured with $V_G$ applied to each gate in (e). (**f**) SCDs measured for the device in (d) at different $V_G$ values (left panel) and different power dissipated by the gate $P_G = V_G \cdot I_{leak}$ (right panel) with opposite polarity of $V_G$ (in both panels, curves for positive $V_G$ are shown in red, while those for negative $V_G$ in blue). Panels from (d) to (f) are adapted from T. Elalaily *et al.*, ACS Nano **17**, 5528-5535 (2023) [31]; licensed under a CC BY license.

In ref. [16], the correlation between $I_{leak}$ and the device $1/f$ noise has been studied, and the results reported show a strong correlation between these parameters. Time-domain measurements carried by the authors show fluctuations between the normal and superconducting state, which have been attributed to filling and emptying of trap states in the oxide along the $I_{leak}$ path (see also section 4.1.2), which occurs via phonon emission. Moreover, at specific $V_G$ (inducing in turn a finite $I_{leak}$) and $I_{bias}$ settings, a resistive state smaller than the normal-state resistance has been observed, which has been attributed to only a part of the device being driven into the normal state.

In ref. [32], Joule heating in a pair of electrodes electrically disconnected from the S wire, results in the suppression of the $I_c$. A similar phenomenology is observed through application of $V_G$ directly to the S wire. These observations are interpreted as the result of decay of high-energy electrons into phonons travelling to the S wire, meaning as a phonon-mediated GCS, which correspond to either scenario 2 or 3. In addition, when the authors cut a trench into the substrate between the gate electrodes and the nanowire, a suppression of the GCS is observed, which also supports the picture of $I_{leak}$-induced phonons. However, the SCD measured under $V_G$ application are much broader than those caused by Joule heating. Therefore, in the sense of the classification used here, this would correspond to scenario 3. Consequently, in Table 2 ref. [32] is assigned to both scenarios 2 and 3.

In a recent study, Zhang and co-workers [33] have also fabricated a gated nanowire connected in series to meandering nanowires, which are typically used for phonon detection. Thanks to the high large kinetic inductance of this device connected in series to a low-noise amplifier and an oscilloscope, the authors have been able to correlate the pulse count, while driving the nanowire into the normal state through an applied $V_G$, to high-energy electrons and phonons excited by $I_{leak}$ in the substrate (Si without an insulating $SiO_2$ layer in this case). Moreover, the authors have showed that the $I_c(V_G)$ characteristics of their devices are asymmetric for opposite $V_G$ polarity, and that this asymmetry can be modulated by varying the sample $T$ because they argue that high-energy electrons are less affected by $T$ variations compared to phonons.

### 3.4. $V_G$-induced mechanism (direct field effect)

Experiments supporting a direct field effect (scenario 4) have also been reported [12,15,17,19-22,25-27,35,46-47]. A first thing to note is that, with the exception of refs. [27,35], these experiments have been carried out in the same lab, although they involve different types of devices, device architectures, S materials, and measurement protocols.

From the evidence reported above from other groups, however, it is clear that an $I_{leak}$ is present in any device and that this always contributes to some extent to the GCS. Therefore, it is challenging to exclude all $I_{leak}$-mediated scenarios and prove a direct field effect.

One of the experiments supporting scenario 4 has been carried out by Rocci and co-workers on gated Ti nanowires which are suspended above the substrate and decoupled from the gate electrodes [22], as shown in Figs. 7a and b. According to the authors of this study [22], the observation of a GCS in this device (Fig. 7c) should rule out any contributions to the GCS due to $I_{leak}$, and therefore scenarios 2 and 3, because the nanowire is completely decoupled (suspended) from the substrate, meaning that no $I_{leak}$-induced phonons or electrons can reach the nanowire from the gate electrode through the substrate.

High-energy electrons, however, can still be injected from the gated into the suspended S nanowire. The authors of this study [22] also exclude this possibility (i.e., scenario 1) based on finite-element simulations. Their simulations show that the current made of high-energy electrons tunneling from the gate into the nanowire across vacuum ($I_{FE}$) at $V_{G,offset}$ (~ 15 V; Fig. 7c) is by several orders of magnitudes lower than that corresponding to $I_{leak}$ (~1.5 nA at $V_{G,offset}$), as shown by the data in Figs. 7d to f. To obtain an $I_{FE}$ comparable to $I_{leak}$, an $E$ of 1÷ 10 GV/m is required according to the calculations done in ref. [22]. Nonetheless, the simulations show that the $E$ at the S surface is at least one order of magnitude lower than the $E$ needed for $I_{FE}$.

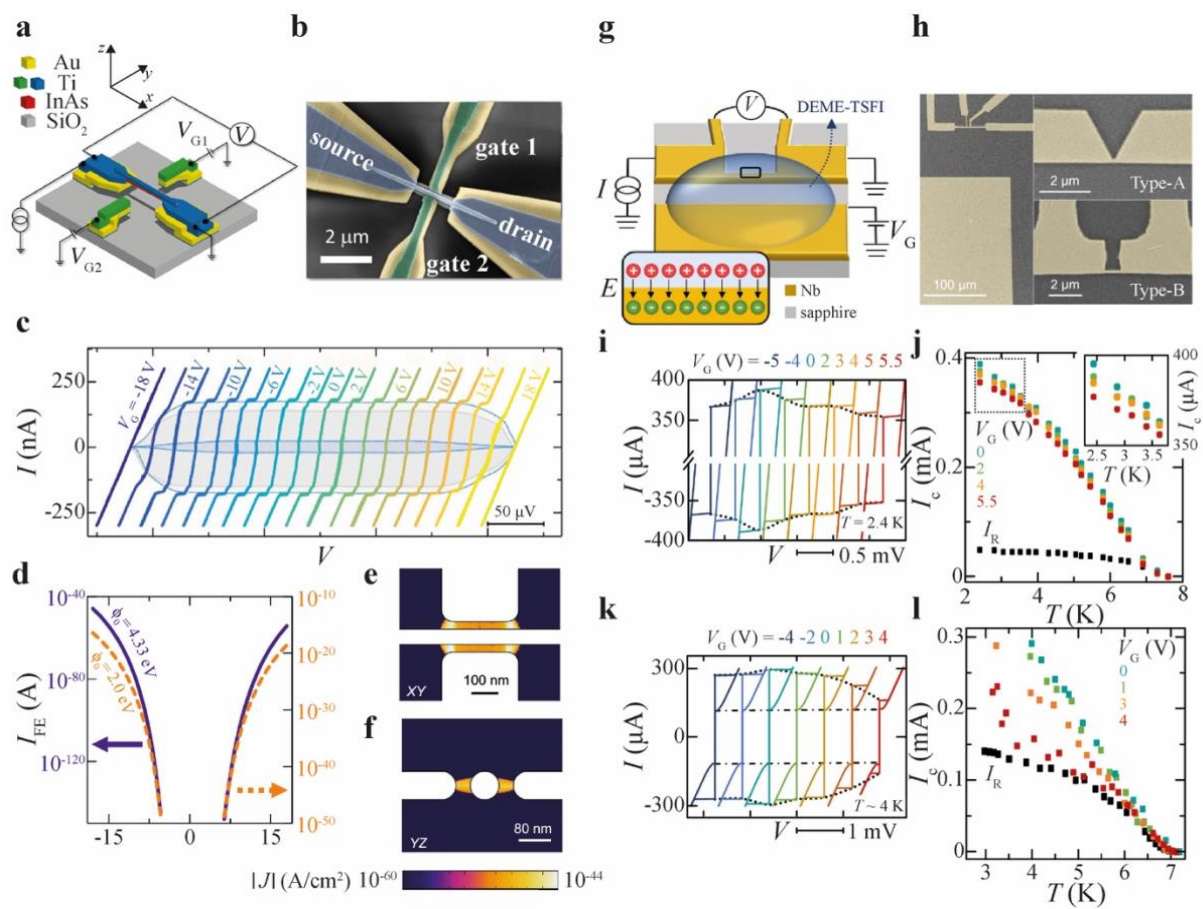

**Figure 7. Evidence for field effect.** (**a**) Schematic of a device consisting of a suspended Ti nanowire with side gates and (**b**) corresponding colored SEM image of an actual device. (**c**) Current versus voltage, $I(V)$, characteristics for the device in (b) at few representative gate voltages $V_G$ showing a suppression of critical current as $V_G$ is increased. (**d-f**) Current due to field emission $I_{FE}$ determined from finite element analysis simulations for the device in (b) assuming the work function $\phi$ of Ti of 4.33 eV (purple curve) and of 2.0 eV (orange curve), and simulated current density $J$ for the same device in the xy-plane (e) and yz-plane (f) according to the reference systems of cartesian axes defined in panel (a). Panels from (a) to (f) are reproduced with permission from M. Rocci *et al.*, ACS Nano **14**, 12621-12628 (2020) [22]. Copyright 2020 American Chemical Society. (**g-h**) Schematic of a device with $V_G$ applied through ionic liquid gating (ILG) and colored SEM images of devices with ILG and made through additive patterning (type A) and through subtractive patterning (type B) in (h). (**i-l**) $I(V)$ curves (i) and $I_c$ versus temperature $T$ dependence at several $V_G$ (as labelled in the legend) in (i) and (j), respectively, for devices of type B. Same data for the devices of type A are shown in (k) and (l), respectively. Panels from (g) to (l) are reproduced with permission from F. Paolucci *et al.*, Nano Lett. **21**, 10309-10314 (2021) [25]. Copyright 2021 American Chemical Society.

In another experiment in support of the direct field effect, $V_G$ has been applied using ionic liquid gating (ILG) [25]. The motivation behind this study is that, if a GCS is observed also with ILG, then the GCS cannot be due to either field emission or any type of $I_{leak}$-induced process. This is because, although ionic liquids have a non-negligible $I_{leak}$ (> 1 nA) already at $V_G$ of few volts [62], movement of charges and therefore $I_{leak}$-induced processes are virtually absent when an ionic liquid get frozen upon cooling and a $V_G$ is applied to it. The authors of ref. [25] indeed observe a GCS with ILG, although the suppression of $I_c$ is not complete (i.e., $I_c$ is not shown to be reduced fully to zero at a certain $V_{G,offset}$), as shown in Figs. 7i to l. In particular, the amount of $I_c$ suppression is different depending on whether the device has been made following a bottom-up approach based on additive patterning or a top-down approach based on subtractive patterning (Figs. 7j and l). This aspect is discussed in more detail in section 4.

We note here that ILG can also induce other mechanical effects due to electrostriction of the liquid [63] or chemical changes in the oxide passivation layer on the S surface [64], which can in turn affect the superconducting properties of the S nanoconstriction [63,64]. A $V_G$-induced modulation in the $I_c$ of Nb micro-bridges has also been reported by other groups using ILG [63], although the authors have also observed a $T_c$ shift of the S (concurrent with a $V_G$-induced modulation in $I_c$), which is usually not observed in GCS-controlled superconducting devices.

Very recently, Yu and co-workers [27] have also reported the GCS in Nb devices with a top-gate geometry. For these devices, $V_{G,offset}$ decreases as the thickness of the dielectric $SiO_2$ layer (used as insulator to decouple the top gate from S) is reduced, meaning when the $E$ strength increases. Also, when other top-gate electrodes are added away from the S nanoconstriction, no GCS is observed, despite the larger $I_{leak}$ measured for these gates compared to the vertical gate. These results and the non-monotonic dependence of $V_{G,offset}$ on $T$ are considered by the authors of ref. [27] as evidence in support of scenario 4 in their devices.

In another experiment, where the GCS is ascribed to an $E$-driven effect, it has also been found that the width of the SCDs increases under an applied $V_G$, which the authors ascribe to action of the $E$ distorting the phase of the superconducting condensate [21]. This broadening is similar to that reported in refs. [31-32,50] (supporting scenario 3) and shown in Fig. 6c.

Experiments carried out on SQUID interferometers also show that the applied $E$ systematically distorts the current-phase relationship of the device, rather than randomizing the phase [47]. The systematic distortion of the phase has been argued by the authors of ref. [47] to constitute evidence against *arbitrary* phase fluctuations (scenario 3). According to the same authors, their measurements suggest that an $E$ can influence the phase of the S nanoconstriction, although no $I_{leak}$ would be at the basis of the mechanism.

Last but not least the two reports [27,37] showing an enhancement in $I_c$ upon application of $V_G$ (in one of them [27], only for a certain temperature range below $T_c$) may also suggest a direct field effect, simply because any $I_{leak}$-induced mechanisms should involve dissipation and hence suppress superconductivity rather than enhancing it. In one of these studies [37], Rocci and co-workers argue that the applied $V_G$ can strongly affect the spin-orbit coupling (SOC) at the surface of the S nanoconstriction, which in turn modifies the vortex surface barrier. Modifications in the vortex surface barrier are considered responsible for the observed enhancement in $I_c$, and consistent with other experimental features like the absence of changes in $T_c$ and the bipolar nature of the observed effect.

To understand which $E$-driven effect can lead to a GCS, several microscopic and phenomenological models based on Ginzburg-Landau theory have also been formulated, which provide good qualitative agreement with experiments [65-72]. Some of these models assume that the $I_c$ suppression is due to a distortion of the superconducting order parameter induced by $E$. More recently, it has also been suggested that magnetic impurities, which can be present in the native surface oxide of a S, can assist pair breaking under the application of an $E$ [72] (see section 3.4.1 for further details).

Finally, a microscopic theory within the Bardeen-Cooper-Schrieffer formalism has also been developed by Zaccone and Fomin [73], which considers the confinement in a S nanoconstriction on the Fermi energy and density of states at the Fermi level. This theoretical model predicts the emergence of a critical $E$ required to suppress superconductivity, which also decreases as the thickness of the S nanoconstriction is reduced, in agreement with experiments.

The Table 3 below summarizes the main pieces of experimental evidence reported to date in support of the different scenarios for the GCS, and which have been discussed in this section.

| **Scenarios proposed** | **Main experimental evidence** |
| --- | --- |
| 1) Field emission | Broadening in the quasiparticle coherence peaks in the device DoS concurrent with $V_G$ application; measurement of $I_c$ suppression in STM setup under tunneling current injection. |
| 2) Phonon heating | Changes in device parameters (e.g., $f_0$ or $Q$ for a resonator) under $V_G$ like those obtained with increase in $T$ and no $V_G$; significant reduction in $V_{G,offset}$ (10% or larger) as device $T$ is increased. |
| 3) Phase fluctuations | Broadening of SCDs under $V_G$ application and analysis of switching dynamics not supported by phase slips only thermally activated; SCDs better matched for same $P_G$ other than for same $V_G$ (with different $V_G$ polarity), but in a way inconsistent with scenario 1. |
| 4) Field effect | Observation of GCS in suspended nanowires (detached from the substrate) and also in devices gated with ionic liquid. |

**Table 3.** Main experimental evidence in support of different scenarios proposed for GCS.

### 3.4.1. *E*-driven effects in metallic superconductors

Before reviewing the experimental parameters affecting the functioning and performance of GCS devices, in this section we describe in more details some of the microscopic models that explain the GCS in metallic superconductors as result of the interplay between superconductivity with an $E$ or gate-induced electrostatic potentials. The main idea underlying each of these models is schematically illustrated in Fig. 8.

The first model reported in ref. [65] addresses the influence of an $E$ on a S as a source of inversion symmetry breaking at the S surface, and it emphasizes the effects of the $E$-induced orbital moments at the S surface on electron pairing. In general, the electronic structure of most superconductors stems from orbital configurations that possess non-zero orbital moments. This is exemplified by the *d*- and *p*-orbital band structures present in elemental Ss made from transition metal elements, and it is relevant for a wide range of materials, including most 2D superconductors, heavy fermions, and superconductors based on iron or chromium, among others.

Recently, it has been recognized that an orbital analog of the spin Rashba effect emerges in acentric crystals or when external fields, such as electric or strain fields, break inversion or mirror symmetry. This resulting orbital Rashba coupling affects the orbital structure of the electronic states by creating orbital moment textures and, remarkably, can occur even without atomic spin-orbit coupling. For multiorbital superconductors, there is an internal degree of freedom associated to the phase of the pairing amplitude for Cooper pairs with a different orbital character.

The investigation of the impact of orbital-dependent acentric interactions reported in ref. [65] indicates that, above a certain critical threshold, the relative phase of Cooper pairs with different orbital character can undergo a transition from 0 to π. Fig. 8a shows the phase reconstruction resulting from the applied *E* for an electronic structure defined by three orbitals, such as *p*- or *d*-orbitals, which belong to an $L = 1$ manifold. This configuration, which is characterized by π-pairing, meaning by an antiphase relationship between superconducting order parameters, causes a sign reversal in the effective Josephson coupling between Cooper pairs, which can in turn lead to a sign reversal of the supercurrent flowing through S. As a result, the authors of ref. [65] show that *E*-driven orbital-phase frustration in an inhomogeneous S represent a viable mechanism for the reduction of $I_c$.

Another *E*-driven mechanism proposed in ref. [67] and that can suppress $I_c$ by acting on the phase coherence of a S involves the generation of vortex-antivortex pairs, characterized by a persistent orbital supercurrent. These orbital vortices can be induced by an *E* or a strain gradient applied at the surface of a S (Fig. 8b). Given that vortex motion contributes to the phase dynamics of a S, the presence of these vortices is expected to lead to dissipative phenomena as S transitions into the normal metal state.

Another microscopic scenario that has been proposed in ref. [68] considers magneto-electric effects due to the supercurrent flow or *E*-driven modification of the magnetic exchange. One form of magneto-electric phenomena in superconductors is typically described by the Edelstein effect, where the flow of supercurrent can generate a finite magnetization, potentially harming the superconducting state or leading to complex phase dynamics (Fig. 8c). In this context, an increase in electrostatic potential can enhance the magnetization created by current flow resulting in values of the magnetization that are sufficiently high to suppress superconductivity. The induced magnetization can be especially pronounced when considering both spin and orbital moments [68].

Another relevant *E*-driven mechanism proposed for the GCS involves the magnetic exchange between magnetic impurities in the surface layer and the spin moments within the S (Fig. 8d). According to the authors of ref. [72], when an *E* is applied, this magnetic exchange becomes activated and intensified, leading to considerable surface depairing through spin-flip scattering processes. As a result, the $I_c$ of the S decreases as the applied *E* is increased [72].

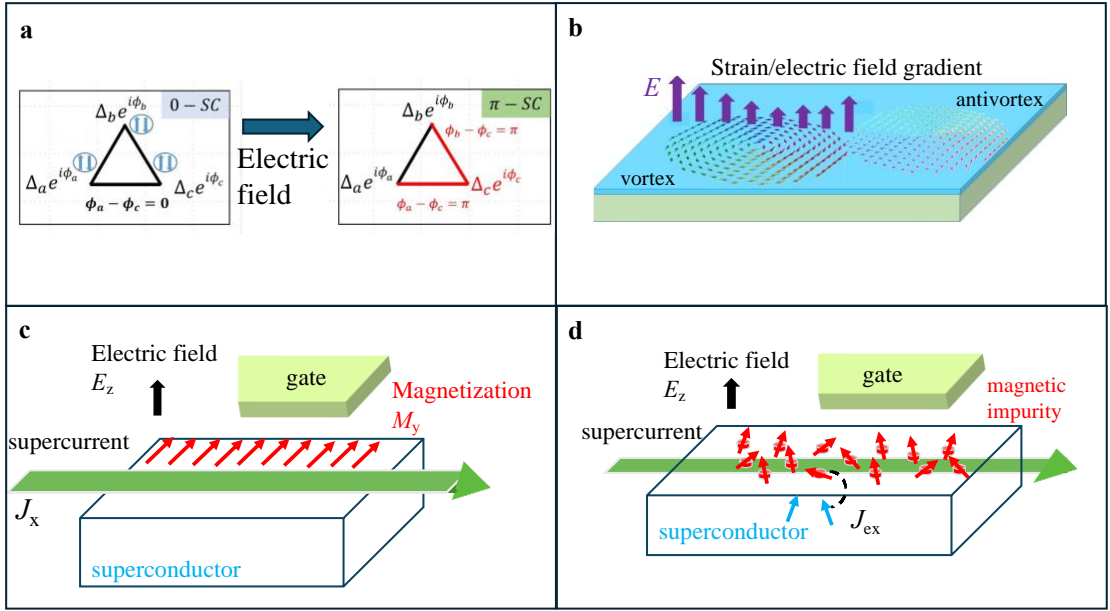

**Figure 8. Physical scenarios and mechanisms for GCS due to an applied electrostatic field**. **(a)** Illustration of the induced orbital antiphase p-pairing, resulting from the inversion-asymmetric interaction caused by an electric field $E$ [65]. The supercurrent suppression arises from the frustration of the superconducting phase, which produces alternating signs in the supercurrent through inhomogeneous weak links within the superconductor S. **(b)** Sketch of the vortex-antivortex pairs generated by an $E$ or strain field that disrupts inversion and mirror symmetry at the surface of a S [67]. The presence of vortices, which give rise to dissipative phase dynamics, is responsible for the suppression of supercurrent. **(c)** Illustration of the magnetization induced by the current flow in the presence of an applied $E$. The applied $E$ induces a non-vanishing magnetization that in turn can be detrimental for the superconducting state or induce non-trivial phase dynamics [68]. **(d)** Schematic of the magneto-electric effects mediated by magnetic impurity at the surface of a S. The interaction between the spin of the impurity and the spin of the electrons in S is enhanced by the applied $E$, leading to considerable depairing through spin-flip scattering processes. As a result, the critical supercurrent decreases with an increasing $E$ [72].

## 4. Experimental parameters affecting the GCS and performance of GCS devices

In this section, we discuss whether there are any specific material, device or fabrication parameters that facilitate the GCS observation and/or improve the performance of GCS devices by, for example, lowering their $V_{G,\text{offset}}$. Lowering $V_{G,\text{offset}}$ is desirable for applications because it would help reduce contributions to $I_{\text{leak}}$ coming from the substrate or from the measurement setup (wire shielding contributes to an increase in $I_{\text{leak}}$ because it becomes less effective at higher $V_G$), and it also would allow easier interfacing of GCS-based logics with CMOS (typically operating at $V_G < 5$ Volts [74]).

In addition, a lower $V_{G,\text{offset}}$ would lead to an increase in the fan out, which is given by the number of devices that can be connected in series to a certain device and controlled by its voltage output $V_{\text{out}}$. The $V_{\text{out}}$ of a GCS device in fact depends on its characteristic voltage at $V_G = 0$ (i.e., $I_{c0}R_N$), as shown by Fig. 1e. A lower $V_{G,\text{offset}}$ would imply that $V_{\text{out}}$ can be more easily fed as input signal to the gate (i.e., used as the $V_G$) of another GCS device connected downstream.

## 4.1. Effects of material parameters, device geometry and fabrication process
### 4.1.1. Influence of S type and structural disorder

The first question that we address in this section is whether the choice of any specific S materials for the fabrication of GCS devices systematically leads to lower $V_{G,offset}$ values. To address this question, since there is a large variation not only in the type of S used, but also in other parameters like $d_{gate}$ across devices made with different Ss, it is better to compare GCS devices not by the absolute $V_G$ needed for $I_c$ suppression, but rather by the $E$ needed to observe such a suppression. Unfortunately, however, $E$ is a parameter not reported in the literature because, unlike $V_G$, it cannot be easily measured experimentally.

In first approximation, however, but without implying that a direct field effect (i.e., scenario 4) is the mechanism underlying the GCS, one could divide the $V_{G,onset}$ and $V_{G,offset}$ values reported (and listed in Table 2) by the $d_{gate}$ of the corresponding devices, to obtain $E_{onset}$ and $E_{offset}$, respectively. Following this approach, only the capacitive coupling between the gate and the S nanoconstriction via vacuum (acting as the dielectric) is considered, while the coupling between the gate and the substrate is neglected. For devices with side gates, which make up almost for all GCS devices studied to date, the determination of $E$ based on such approach should not lead to significant errors because $E$ in the substrate is reduced by its relative permittivity $\varepsilon_r$, which implies that the $E$ component in the substrate is significantly smaller than the $E$ component through vacuum. The validity of this approach is evidenced by the fact that the estimates obtained for $E$ are consistent with those calculated based on more sophisticated tools like finite-element method simulations [75].

For devices with a top-gate or back-gate geometry, where $SiO_2$ has been used as dielectric [12,27], $E$ can be estimated by dividing $V_G$ first by $d_{gate}$ and then by the $\varepsilon_r \sim 4$ of $SiO_2$ at low $T$ [76]. Similarly, for refs. [14,28], where hexagonal boron nitride (hBN) has been used as dielectric, $V_G$ can be scaled by $d_{gate}$ and $\varepsilon_r \sim 3$ of hBN [77] to estimate the corresponding $E$. For ref. [25] where ILG has been used, however, this procedure cannot be followed because, if $V_G$ were divided by $d_{gate}$ ($\sim 10^5$ nm in ref. [25]), an unrealistically small $E$ would be obtained. To get a better estimate of the actual $E$ in the case of ILG, the thickness of the electronic double layer forming at the interface between the liquid and the S nanoconstriction, as well as its charge distribution, should be known. Although in ref. [25] the actual thickness of the electronic double layer is not reported, based on other studies [63,78], we estimate that $E$ varies between 10 and 100 MV/cm for a $V_G$ of few volts applied in ref. [25]. This value is consistent with the magnitude of $E$ in other GCS studies without ionic liquid gating, where $E$ has been estimated using the procedure described above.

Fig. 9a shows the $E_{onset}$ and $E_{offset}$ values for GCS devices obtained based on the considerations listed above. The $E$ values have been grouped in Fig. 9a according to the S material used in their corresponding devices, with the S materials arranged to have increasing atomic number $Z$ along the positive direction of the horizontal axis of Fig. 9a.

For a given S, Fig. 9a shows that there exists a large variation in $E_{onset}$ and $E_{offset}$ values. This variation is most likely due to the fact that, with the exception of ref. [29], where a statistically relevant number of GCS devices made with the same S (Nb) have been studied, in all the other studies carried to date on the GCS, only a few devices (typically one or two) have been characterized in each study. As a result, for a specific S material, the data points in Fig. 9a refer to devices made by different groups, where other parameters such as the device geometry and $I_{leak}$ differ significantly.

The only trend that can be inferred from Fig. 9a is that $E_{onset}$ and $E_{offset}$ tend to decrease in Ss with higher $Z$. The trend is possibly even clearer in Fig. S2a which shows the same data as Fig. 9a but on a linear scale. Since physical parameters like SOC increase with $Z$, and since SOC is considered as a relevant parameter in models proposed to explain the GCS [65,70], the SOC strength of a S can be important to reduce $E_{offset}$.

More systematic studies, however, are necessary to verify the existence of a correlation between $E_{offset}$ and the SOC strength, which may also give further insights into the physics of the GCS. Other properties related to $Z$ include, for example, the complexity of the Fermi surface (e.g., number and topology of electronic bands, symmetry of the electron-phonon coupling strength). Their role toward the GCS also remains to be explored.

**4.1.2. Influence of substrate material**

Fig. 9a shows that an average $E_{offset}$ of ~ 3-4 MV/cm is needed for a full $I_c$ suppression in most devices fabricated on insulating substrates like $SiO_2$ or $Al_2O_3$ [12,19-20,22]. This value gets significantly reduced when GCS devices are made on non-insulating substrates like Si, for which $E_{offset}$ is typically below 1 MV/cm [23,32] (blue data points in Fig. 9a). Devices made on Si also show asymmetric $I_c(V_G)$ curves, for which $V_{G,offset}$ is different depending on $V_G$ polarity, as shown by the $I_c(V_G)$ data in ref. [23]. These two observations suggest that the GCS in devices made on insulating substrates and those made on Si can be governed by different mechanisms. It is very likely that the strong thermal coupling between S and the gate in devices on Si, for example, can cause phonon-induced heating (scenario 2) or field emission (scenario 1). Phonons can also have different effects in $SiO_2$- and Si-based devices, since the average phonon propagation length is of few microns in Si and of ~ 5 nm in $SiO_2$ at 4.2 K [79,80].

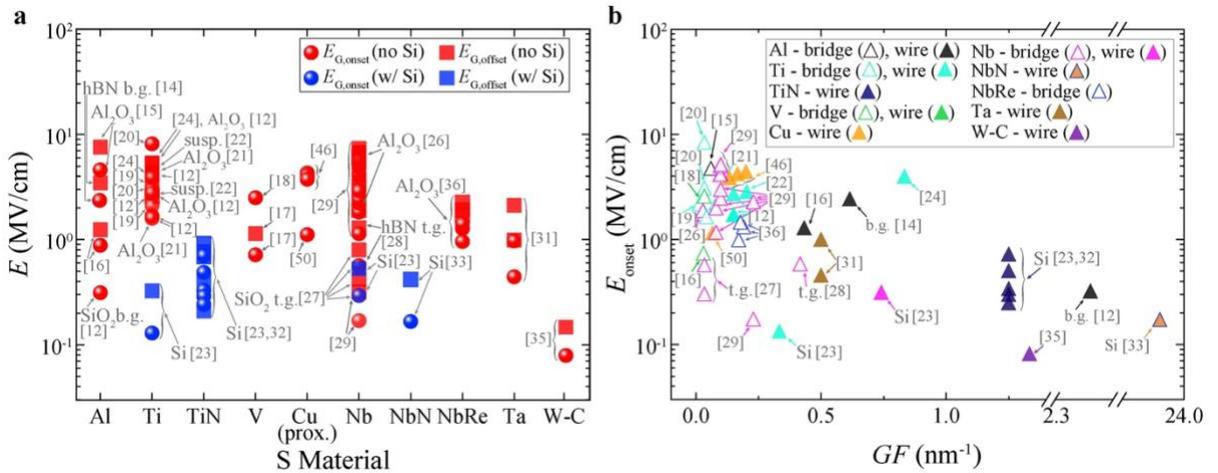

**Figure 9. Dependence of the GCS on device material and geometry.** (**a**) Electric field at 10% of $I_c$ suppression ($E_{onset}$; round symbols) and at full suppression of $I_c$ ($E_{offset}$; square symbols) for different devices as a function of the S material. Blue and red symbols are used for devices made or not made on Si, respectively. (**b**) $E_{onset}$ as a function of the geometry factor $GF = l_S/(w_S \cdot t_S)$ (with $l_S$ = length, $w_S$ = width and $t_S$ = thickness of the S constriction) for different gate-controlled superconducting devices made of different S materials (specified in the panel legend), with hollow and filled symbols used for Dayem bridges and nanowires, respectively. In both panels, the reference number is indicated next to corresponding datum point, and the acronyms t.g. and b.g. stand for top gate and back gate, respectively.

Recent studies [16,29] have shown that devices made on substrates like $SiO_2$, which are prone to exhibit stress-induced leakage current (SILC) effects due to oxygen migration under the relatively high $E$ applied in GCS devices [81,82], can show a change in their working point (e.g., their $V_{G,offset}$) over time. This is shown by Fig. 10, which reports data from a recent study by Ruf and co-workers [29], where it has been found that, after the $SiO_2$ substrate experiences an increasing $I_{leak}$ (Fig. 10a), the GCS device can suddenly jump to another working point, which is characterized by a reduction in the $V_{G,offset}$ of its $I_c(V_G)$ characteristics (up to ~ 20 V in ref. [29]). Although a SILC event can lead to a significant reduction in $V_{G,offset}$, however, the $I_c$ suppression (normalized by its initial value $I_{c0}$) always follows the same dependence on the power dissipated by the gate $P_G$, as shown by Fig. 10c. This is because, although $V_G$ decreases after a SILC event, the $I_{leak}(V_G)$ curve shifts toward higher $I_{leak}$ values (at fixed $V_G$) as shown in Fig. 10b, due to the formation of more conducting channels in the $SiO_2$ substrate (Fig. 10b), which makes the product $I_{leak}V_G = P_G$ constant.

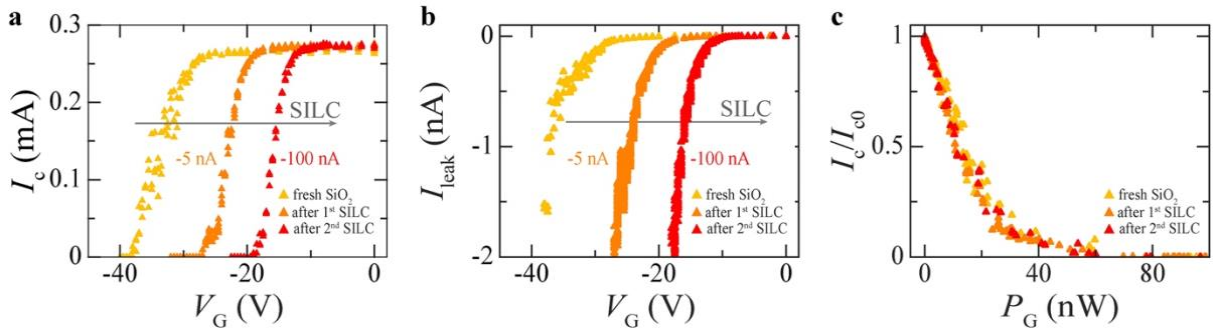

**Figure 10. Effect of stress-induced leakage current in the substrate on GCS devices.** (**a-b**) Critical current $I_c$ versus gate voltage $V_G$, $I_c(V_G)$, curves in (a) and leakage current $I_{leak}$ versus $V_G$, $I_{leak}(V_G)$, curves in (b) measured after inducing subsequent SILC events via the injection of current between the gate and the S constriction (the current values are specified next to each curve). (c) Dependence of $I_c$ normalized to its value at $V_G = 0$ ($I_{c0}$) on power dissipated by the gate $P_G$ after each SILC event. All panels are from L. Ruf *et al.*, ACS Nano **18**, 20600 (2024) [29]; licensed under a CC BY license.

In addition to SILC events, variable stress-induced leakage current (V-SILC) events can also occur because of switchable defects [83-84] located, for example, in $SiO_2$ area of the device placed between the gate electrode and the S constriction. Unlike SILC events, which lead to a stable shift in the working point of the device, V-SILC events can induce instabilities over short timescales and manifest, for example, as fluctuations in $I_{leak}$ (under an applied $V_G > V_{G,onset}$) that are concurrent with fluctuations in $I_c$. This strong correlation between noise in $I_{leak}$ (i.e., fluctuations in $I_{leak}$) and voltage fluctuations or fluctuations in $I_c$ of the S constriction has been measured by two different groups [16,29], on both short timescales and long timescales (Fig. 11). Both groups have also interpreted their results as consistent with scenario 3.

Although SILC effects can be exploited as a viable approach to pre-train a certain GCS device made on $SiO_2$ and achieve a reduction of its operational $V_{G,offset}$, they also suggest that, for technological applications where strong device stability is required over time, substrates different from $SiO_2$ and less prone to SILC events should be used.

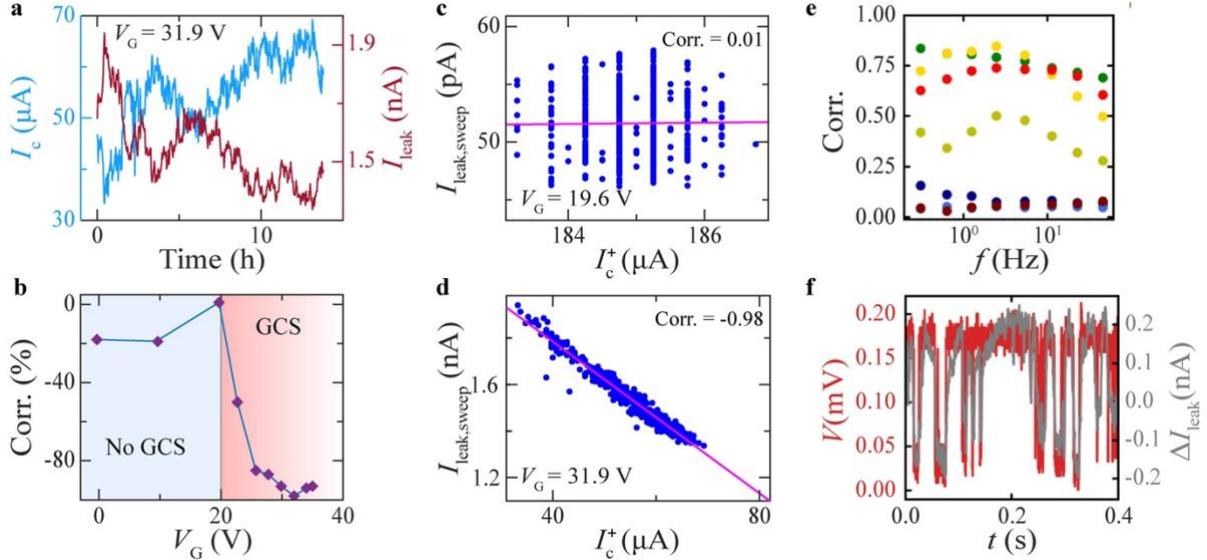

**Figure 11. Effects of variable stress-induced leakage current events in the substrate on GCS devices.** (**a-b**). Evolution of critical current $I_c$ (light blue curve) and leakage current $I_{leak}$ (red curve) over long time scales at an applied gate voltage $V_G \sim 31.9$ V $> V_{G,onset}$ for a Nb GCS device (a) showing that fluctuations in $I_{leak}$ due to variable stress-induced leakage current (V-SILC) events anticorrelate with fluctuations in $I_c$. (b) Correlation factor for the same device as in (a) plotted at a few representative $V_G$ and showing that the anticorrelation increases (in amplitude) when $V_G > V_{G,onset}$ and the GCS effect kicks in. (**c-d**) Average $I_{leak}$, $I_{leak,sweep}$, measured for the same Nb device as in (a) and (b) during an $I$-$V$ sweep while upsweeping the bias current $I$ plotted as a function of the positive $I_c$, $I_c^+$, extracted from the $I(V)$ characteristic for a $V_G < V_{G,onset}$ showing no correlation (c), and for $V_G > V_{G,onset}$ showing almost perfect anticorrelation between $I_{leak,sweep}$ and $I_c^+$. Panels from (a) to (d) are from L. Ruf *et al.*, ACS Nano **18**, 20600 (2024) [29]; licensed under a CC BY license. (**e-f**) Correlation between $I_{leak}$ noise spectrum and noise spectrum in the voltage drop measured across an Al/InAs core-shell nanowire with an applied $V_G$ = 5 V as a function of spectrum frequency in (e) and time evolution of the voltage measured across the nanowire (red curve) and of the variation in the leakage current (gray curve) for the same device at $V_G$ = 5 V and for a bias current injected through the nanowire $I_{bias}$ = 5.3 μA (f). The different colors in (e) corresponds to different values of $I_{bias}$ injected through the nanowire. Panels (e) and (f) are from T. Elalaily *et al.*, arXiv pre-print at https://arxiv.org/pdf/2312.15453.pdf (2024) [16]; licensed under a CC BY license.

### 4.1.3. Influence of device geometry

Studies with systematic variation of parameters related to the device geometry like $d_{gate}$ or and length of the S constriction for the same S material are sparse. Only recently, Ruf and co-workers have carried out a systematic study [29] of a series of GCS devices made of Nb, where all the geometry parameters have been kept fixed except for the width $w_S$ of the S constriction, to study the effect of $w_S$ on the GCS. In addition to showing that the GCS can be also observed for devices with $w_S$ up to 550 nm, and therefore much wider than $\xi_S$ (typically < 15 nm for Nb [85] in the diffusive regime), the authors have also shown that no increase in $V_{G,offset}$ is observed as $w_S$ is increased. These results suggest that side-gated devices with wider S constrictions perform equally well, in terms of $V_{G,offset}$, compared to devices with a narrower constriction, while offering the advantage of being more robust over prolonged thermal cycling and continuous operation.

The large $w_S$ of the gated Nb devices studied in ref. [29] results in a higher $I_{c0}$ and hence in a higher $I_{c0}R_N$ (~ 0.25 V at 1.5 K) compared to gated Nb devices reported by other groups, for which the smaller $w_S$ (< 200 nm) leads to $I_{c0}R_N$ of few tens of mV (see Table in the Supporting Information). GCS devices with large $w_S$ appear therefore promising to increase the fan out in GCS-based superconducting logics. If a different S with higher resistivity and/or critical current other than Nb (e.g., NbN or NbRe) or a longer S constriction were made, the characteristic voltage $I_{c0}R_N$ achieved in ref. [29] could be easily increased to few Volts, which would already allow interfacing of GCS devices with CMOS devices.

Apart from the independence of $V_{G,offset}$ and $E_{offset}$ on the $w_S$ of the device, no other conclusions can be made regarding the effects of other geometrical device parameters on the GCS. To better visualize this, in Fig. 9b we show how $E_{onset}$ varies as a function of a geometry factor (GF) defined as $GF = l_S/(w_S \cdot t_S)$, where $l_S$, $w_S$ and $t_S$ are the length, width, and thickness of the gate-controlled S nanoconstriction, respectively. The same data of Fig. 9b are also shown in Fig. S2b in a linear-linear plot. Devices with larger GF values (> 0.13) are mostly nanowires, whereas lower GF values correspond to Dayem bridges.

The data in Fig. 9b show that Ti devices have similar $E_{onset}$ values, independently on GF. Also, devices made of Ta [31] and W-C [35] exhibit lower $E_{onset}$ (< 1 MV/cm) despite having different gate electrodes (Table 2) and GFs differing by more than 2. Nonetheless, devices made with Al [12,14] show a reduction in $E_{onset}$ by almost one order of magnitude as the GF is increased by a factor of 4.

If the S material is not considered, Fig. 9b suggests a decreasing trend of $E_{onset}$ with increasing GF. Nonetheless, it should be noted that most of the studies performed on devices with higher GF (i.e., longer and/or thinner wire based on the GF definition) are also carried on Si substrates, which per se has lower $E_{onset}$ and $E_{offset}$ values compared to substrates with an insulating layer, as already shown in Fig. 9a.

More systematic studies like those reported in ref. [29] are therefore needed also for other geometry parameters to determine whether they play a role on the GCS or not, since the analysis based on existing studies is not conclusive.

After the first studies on the GCS were published, it was also argued that, in devices with sharp edges, current-crowding effects may appear [86-87], which can in turn affect the GCS. Recently, however, it has been demonstrated that current-crowding has little effects on the GCS. This has been shown in ref. [29], where the authors have characterized devices with a certain $w_S$ and twin devices, where $w_S$ has been reduced (after fabrication) by introducing a sharp edge inside the S constriction with focused ion beam (FIB). The measurements performed do not show any reductions in $V_{G,offset}$ in the devices after the FIB cut, as one would instead expect if current-crowding played a role.

### 4.1.3. Influence of fabrication process

Very recently, it has been shown that the fabrication process followed to make a three-terminal superconducting device is also crucial for the GCS observation. As shown in Fig. 12, devices made following an additive approach involving EBL patterning, deposition of the S material and lift-off (here called lift-off devices) show a GCS, unlike devices made with the same S and geometry but using EBL patterning through a negative resist and etching through the resist mask (here called etched devices). A microstructural analysis of both types of devices made in ref. [34] suggests that the larger roughness and microstrain in lift-off devices compared to etched devices, together with other surface modifications induced by the fabrication process, can account for the absence of the GCS in the etched devices.

The key role played by the fabrication process for the GCS observation has been confirmed by Koch and co-workers [36] who have not only demonstrated that in devices made from highly-disordered NbRe (S) films through subtractive patterning, it is possible to observe the GCS, but also that, in the same devices, the GCS is only observed when the etching process is carried with a specific ($Ar/Cl_2$) gas mixture. When a different gas mixture (e.g., Ar or $Ar/SF_6$) is used, then no GCS can be observed, despite the presence of disorder in the starting S material. These finding imply that the combination of disorder and surface modification induced by the fabrication process plays a role toward the GCS, and are consistent with arguments made in ref. [50] that $I_{leak}$ should flow through surface state for the GCS to occur.

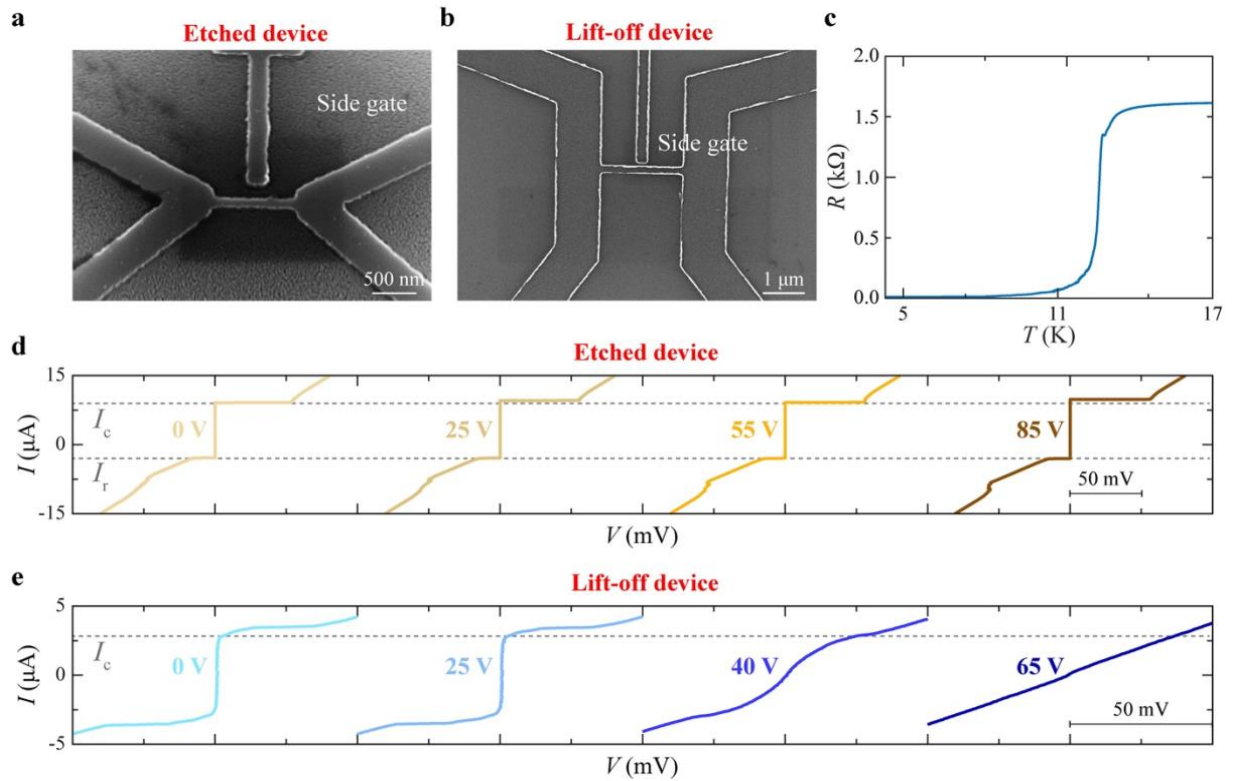

**Figure 12. Effect of fabrication process on GCS.** (**a-b**) Scanning electron microscope images of a NbTiN nanowire devices made by dry etching (a) and by lift-off (b) on a $SiO_2$ (300 nm)/p-doped Si substrate. (**c**) Resistance versus temperature, $R(T)$ curve close around the superconducting transition for the device shown in (a). (**d-e**) Current versus voltage, $I(V)$, characteristics measured for increasing bias current $I$ for the NbTiN device in (a) are shown in panel (d), and $I(V)$ characteristics for the NbTiN

device in (b) are shown in panel (e) for a few representative applied $V_G$ values (indicated next to the corresponding $I(V)$ curve). The data in (d) for the etched device do not show a progressive suppression of either the critical current ($I_c$) or retrapping current ($I_r$) with increasing $V_G$, while $I_c$ is instead suppressed for the lift-off device in (e). All panels are from L. Ruf *et al.*, APL Mater. **11**, 091113 (2023) [34]; licensed under a CC BY license.

### 4.2. $I_{leak}$-induced effects toward prevalent mechanism

In addition to how the GCS depends on the S material, substrate, device geometry, and fabrication route, it is worth considering if any of the mechanisms proposed for the GCS becomes dominant, depending on the relative contribution of $I_{leak}$ to the $I_c$ suppression. One possibility to estimate this contribution is by calculating the ratio $P_{G,offset}/P_N$, where $P_N = R_N I_{c0}^2$ as explained in section 3.2. Fig. 13a shows the ratio $P_{G,offset}/P_N$ for the devices reported in the literature, where this ratio can be calculated based on the data reported, as a function of the S material used in these same devices.

One may argue that, if $P_{G,offset}/P_N \gg 1$ (a ration equal to 1 is marked by a dashed line in Fig. 13a), then $I_{leak}$-induced heating (i.e., scenarios 1 or 2) should represent the main contribution toward the GCS, since $V_{G,offset}$ induces a power dissipation larger than what the device can dissipate after switching to the normal state. Nonetheless, in addition to devices categorized under either of these scenarios (e.g., refs. [14,24]), also devices for which the GCS has been ascribed to other mechanisms like direct field effect (e.g., suspended nanowires in ref. [22]) or phase fluctuations (e.g., core/shell nanowires in ref. [16]) fall within the region of $P_{G,offset}/P_N \gg 1$ in Fig. 13b.

If the $P_{G,offset}/P_N$ ratio is much smaller than 1, it would be difficult to identify a priory a specific mechanism responsible for the GCS, and it is even possible that several mechanisms are at play simultaneously. For $P_{G,offset}/P_N \ll 10^{-3}$, Fig. 13a indeed shows that we do not only find studies where a direct field effect has been proposed [12], but also studies on devices made on Si, where $I_{leak}$-induced effects can be predominant due to the stronger substrate-mediated coupling between gate and S [23,32].

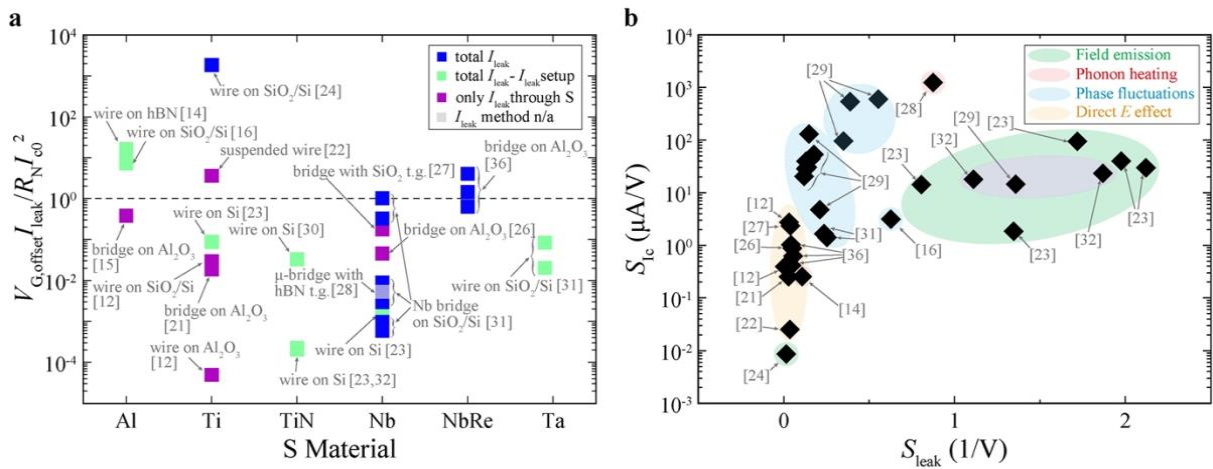

**Figure 13. Effect of dissipated power on the GCS.** (**a**) Ratio between power dissipated by the gate at $V_{G,offset}$ and power dissipated by the device in the normal state, $V_{G,offset}I_{leak}/R_N I_{c0}^2$ as a function of the S material used for different $I_{leak}$ measurement setups (specified in the panel legend). The substrate and device geometry are indicated next to each datum point (t.g. stands for top gate). (**b**) Slope $S_{Ic}$ of the $I_c(V_G)$ curve as a function of the slope $S_{leak}$ of the $I_{leak}(V_G)$ curve for different studies labelled with colored bubbles based on the mechanism proposed to explain the GCS therein. In both panels, the reference number of each study is indicated next to the corresponding datum point.

The difficulty in determining the prevailing mechanism based purely on the $P_{G,offset}/P_N$ value is also due to the discrepancy in the protocols followed for the measurement of $I_{leak}$. In Fig. 13a, the data points have also been differentiated (using different colors) based on the approach followed by the authors of the corresponding study to measure $I_{leak}$. As also discussed in the Supplementary Information, while some groups measure the total $I_{leak}$ between the gate and the device, which also includes contributions from the wiring, others just subtract the contribution to the total $I_{leak}$ coming from the setup [14,23,31-32], and others measure $I_{leak}$ going through a reference resistor placed in series between the device and the electrical ground (Fig. S3) [12,15,21-22,26]. The last method might be the most accurate since it excludes wiring contribution, and consequently it yields lower $I_{leak}$ values (and lower $P_{G,offset}/P_N$ ratios) compared to the two other approaches. Nonetheless, it is restricted to measurements without current bias through the S wire. A standardization in the $I_{leak}$ determination should be therefore introduced in future studies on the GCS to compare devices measured by different groups based on the absolute $I_{leak}$ and $P_{G,offset}/P_N$ values.

Another interesting observation made by different groups [23,29] is that, while $P_{G,offset}/P_N$ can vary significantly for devices with different parameters (e.g., different $V_{G,offset}$) but measured with the exact same setup, the ratio $P_{G,offset}/P_r$ (with $P_r = R_N I_{r0}^2$) remains constant, independently on parameters like $V_{G,offset}$. This result is shown in Fig. 14, which is reproduced from ref. [29]. Here, after studying several Nb devices differing only for their $w_S$, it has been demonstrated that a change in $w_S$ usually corresponds to a change in $V_{G,offset}$ and in the $P_{G,offset}/P_N$ ratio, with the latter getting smaller as $w_S$ increases because a larger $w_S$ in turn leads to an increase in $I_{c0}^2$. Nonetheless, if the ratio $P_{G,offset}/P_r$ is calculated for the same devices, this ratio seems independent on parameters like the $w_S$ and $V_{G,offset}$ of the devices. Whether this observation can give further hints into the mechanism responsible for the GCS remains to be understood in the near future.

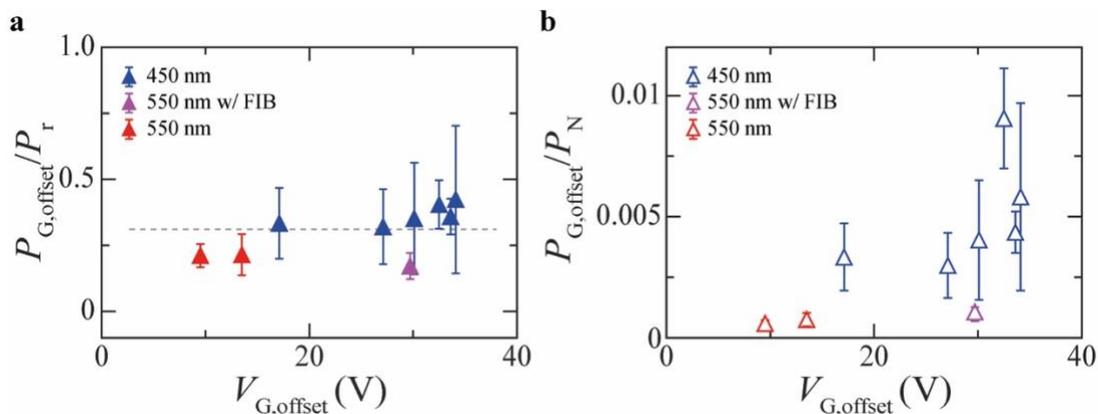

**Figure 14. Dependence of $P_{G,offset}$ in GCS devices on other parameters.** (**a-b**) Power dissipated by the gate $P_{G,offset}$ at gate voltage needed for full suppression of the critical current ($V_{G,offset}$) normalized to $P_r = R_N I_{r0}^2$ in (a) and to $P_N = R_N I_{c0}^2$ in (b) as a function of $V_{G,offset}$ for devices with different widths $w_S$ (specified in the legends of each panel). All panels are from L. Ruf *et al.*, ACS Nano **18**, 20600 (2024) [29]; licensed under a CC BY license.

Since it is difficult to determine whether a mechanism proposed for the GCS becomes prevalent within a certain $P_{G,offset}$ range because $P_{G,offset}$ values are affected by the approach used to measure $I_{leak}$ (Fig. 13a), to carry out such analysis it is necessary to define a parameter that is independent on the absolute $I_{leak}$. To this aim, in Fig. 13b we have defined the slope $S_{leak} = (\text{Log } (I_{leak,offset}/I_{leak,onset}))/(V_{G,offset} - V_{G,onset})$ of the $I_{leak}(V_G)$ characteristics (on a log-linear plot), which contains information on the functional dependence of the $I_{leak}(V_G)$ curves. The reasoning behind is that, depending on the mechanism behind the GCS, a different $V_G$-dependence of $I_{leak}(V_G)$ characteristics can be expected. For example, for field emission of high-energy electrons from the gate to the nanoconstriction, one might expect that the $I_{leak}(V_G)$ trends follows an exponential increase or a power low with a large exponent ($S_{leak} > 1$) [53], while thermal emission would result in a weaker power law dependence. This consideration is supported by the data in Fig. 13b, where indeed we find that $S_{leak} > 1$ corresponds to devices where field emission has been proposed as GCS mechanism.

The power dependence of the $I_{leak}(V_G)$ curves given by $S_{leak}$ can be correlated to the steepness of the $I_c(V_G)$ curves, which we define through another parameter $S_{Ic} = (I_{c,onset} - I_{c,offset})/(V_{G,offset} - V_{G,onset})$. The idea behind our argument is that a specific mechanism is at play, which can be identified by $S_{leak}$ values falling within a certain range, then the same mechanism can also affect $S_{Ic}$, meaning how rapidly the $I_c$ suppression occurs under an increasing applied $V_G$.

In Fig. 13b we therefore show $S_{Ic}$ versus $S_{leak}$ with the $S_{Ic}$ axis on a log scale, to determine 1) if a higher power law meaning a faster rise in $I_{leak}$ (i.e., a higher $S_{leak}$) always correlates with a higher $S_{Ic}$ meaning with a steeper decay in $I_c(V_G)$ and 2) if any specific mechanisms suggested for the GCS always occurs within specific values of $S_{Ic}$ and $S_{leak}$.

As Fig. 13b shows, the studies to date for which $S_{Ic}$ and $S_{leak}$ can be calculated mostly fall into two groups, one with $S_{leak} \sim 0.1/V$ or lower, and the other one with $S_{leak} > 0.1/V$, suggesting two distinct populations. In the first regime, although the $S_{Ic}$ values are scattered over a broad range (i.e., over five order of magnitudes), they mostly remain smaller than 10 µA/V (i.e., the suppression of $I_c$ with increasing $V_G$ is slower). Studies for which a direct field effect has been suggested mostly fall within this regime [12,26,31], meaning this mechanism goes along with a weak increase of $I_{leak}$.

The experiments suggesting phase fluctuations, phonon heating or field emission, fall into the second case, meaning $S_{leak} > 0.1/V$. In particular, studies for which phonon heating and field emission have been proposed, have $S_{leak} > 0.8/V$. In this range, $S_{Ic}$ also appears to be independent of $S_{leak}$ and it adopts values over a wide range (from intermediate to large values), where the $I_c$ suppression with $V_G$ is not very slow. Devices falling in this regime include those made directly on Si [23,32] where, although the absolute $I_{leak}$ is relatively small, the increase of $I_{leak}$ with $V_G$ (i.e., $S_{leak}$) is also large. The conclusion drawn from this analysis so far is that the steepness of the $I_c$ suppression with $V_G$, meaning $S_{Ic}$, does not give immediate information on the mechanism underlying the GCS, since it spans over wide ranges, whereas the mechanism can be more easily inferred from the correlation between $I_{leak}$ and $V_G$, meaning based on $S_{leak}$.

# 5. Technological applications based on the GCS

Understanding the mechanism behind the GCS and the material and device parameters that are key to control it are necessary steps to develop any technological applications based on the effect. For instance, any mechanism including heating and subsequent relaxation might limit the speed and the maximum integration density of GCS devices. The development of technologies like GCS-based superconducting logics also requires overcoming other challenges that are common for integrated circuits. These challenges include finding S materials that allow a reduction in $V_{G,offset}$ and fabrication protocols that ensure high reproducibility in the observation of a GCS, increasing the number of devices that can be controlled downstream by the $V_{out}$ of a given device (i.e., the fan out [88-89]), testing the highest switching speed of GCS devices $f_{max}$ and realize more complex circuits based on them. If these challenges are overcome, competitive GCS-based superconducting logics and other technological applications based on the GCS can be developed.

## 5.1. GCS for superconducting logics

GCS logics can have substantial advantages over CMOS logics and state-of-the-art superconducting logics like rapid single flux quantum (RSFQ) logic [8,90-91]. This would particularly hold true if scenarios 3 or 4, i.e., mechanisms with small dissipation, would be responsible for the GCS.

Table 4 shows a comparison between performance parameters of different technologies for logics, based on the assumption that scenario 4 was the mechanism underlying the GCS. First, GCS-based superconducting devices are easier to scale up compared to RSFQ devices. This is because RSFQ devices have larger dimensions than GCS devices because they are controlled via an $I_{bias}$ (or via an applied magnetic flux). Based on the dimension of GCS devices reported and considering the space for load resistors, GCS-based logics can have a density up to three orders of magnitude higher (~ 10 devices/$\mu m^2$) than RSFQ (~ 4·$10^{-2}$ devices/$\mu m^2$; ref. [91]).

In addition, if top-gate contacts other than side contacts were systematically adopted the $V_G$ application as done in ref. [27], then an even higher device density (> 25 devices/$\mu m^2$) could be achieved, which is comparable to that of CMOS [92]. We note that, with the exception of ref. [27], the application of $V_G$ through top gates has not been systematically tested in GCS devices, most likely because the growth of an insulating barrier on top of a S without pin holes and high breakdown voltage is challenging. Top-gate devices with thin insulating layers, however, are a very promising route to explore also to achieve a reduction of $V_{G,offset}$.

| Performance parameter | Technology for logics | | |
|---|---|---|---|
| | **GCS logics** | **RSFQ** | **CMOS** |
| **Switching mechanism** | Under investigation | Magnetic Flux | Field Effect |
| **Switching energy $E_S$ (J/flops)** | $10^{-21} \div 10^{-19}$ [8] | $10^{-19}$ [6,100] | $10^{-12}$ [100] |
| **Density (#devices/$\mu m^2$)** | > 10 [12,21] | 4 x $10^{-2}$ [91] | ~$10^2$ [*] |

| Switching speed (GHz) | $10^3$ [99] | $7.7 \times 10^2$ [87] | 5 [93] |
|---|---|---|---|
| **Robust against magnetic fields** | YES [20, 45] | NO [96] | YES |
| **Fan-out** | Virtually unlimited (lithographically limited) | 1÷3 [101] | Virtually unlimited (lithographically limited) |
| **Gate-to-channel resistance ($\Omega$)** | $10^{12} \div 10^{13}$ (at 4.2 K) [26] | n/a | $\sim 10^{12}$ (at 4.2 K) |

*values estimated

**Table 4 – Comparison between performance of GCS logics with RSFQ and CMOS logics.**

Second, GCS logics can be faster than CMOS technology. Although the highest $f_{max}$ for CMOS can be above 100 GHz [93], $f_{max}$ is usually limited to ∼ 5 GHz in CMOS high-performance computing systems [94] to avoid overheating. Although the $f_{max}$ of GCS logics has yet to be measured and can be affected, for example, by the mechanisms underlying the GCS, $f_{max}$ can be in principle limited only by the frequency of the S gap $f_G = 2\Delta/h$ (*h* being the Planck constant) – which can be > 1 THz for Ss like NbTiN [95]. We note that superconducting logics like RSFQ have already been successfully driven up to ∼ 770 GHz [96].

Even if the GCS is driven by an $I_{leak}$-triggered mechanism, which can limit the highest $f_{max}$ achievable, as argued in ref. [16], using superconductors with high $T_c$, the operational temperature of the device can be increased to several K, where electron thermal cooling will be replaced by electron-phonon heat transfer [97]. This can also increase the $f_{max}$ to the hundreds of GHz range, which is suitable for the realization of ultrafast superconducting logics [16].

Like for RSFQ, the other main advantage of GCS-based logic over CMOS is its lower power dissipation. Although a $V_G$ of few tens of Volts is typically required to control a GCS device, a pre-bias $|V_{pb}| > V_{G,onset}$ can also be applied to the entire circuit (e.g., via back gating) such that each device can be then controlled only with a small local $V_G$ applied (e.g., via a side gate) on top of $V_{pb}$. Defining the switching energy as $E_S = \frac{1}{2} L(I_c)^2$ [98-99], and using typical values of $I_c \sim 10 \div 100$ μA and $L \sim 10 \div 100$ pH [12,26], one gets $E_S \sim 10^{-21} \div 10^{-19}$ J/flop for a GCS device, which is similar to RSFQ and significantly lower than CMOS [6,100].

Unlike RSFQ devices, GCS devices, independent of the mechanism at play, are also robust against environmental magnetic noise and have good decoupling between input and output signals thanks to their three-terminal geometry with a gate-to-channel resistance ∼ 1-10 TΩ at 4.2 K similar to CMOS at the same *T* [101-102]. The decoupling is essential for high directionality in the transmission of signals and to reduce cross-talking between neighboring cells. Also, RSFQ devices usually have a low fan out (between 1 and 3; ref. [102]), while a single GCS device can be used to drive more devices connected to its output, provided that $V_{out} > V_{G,offset}$ for the devices downstream. This requirement would be easier to meet if the typical $V_{G,offset}$ for the GCS gets reduced.

The other main advantage of GCS-based logics is that, unlike RSFQ, it does not need an interface layer to be connected to CMOS. Although several ways to realize the interface layers between RSFQ and CMOS are currently being tested [103], GCS logics is naturally compatible with CMOS without any additional interfaces, since both technologies are $V_G$-controlled.

We note that other types of materials and devices have also been proposed or are under study for the development of superconducting logics as an alternative to RSFQ – which remains the only one commercially available. An alternative to a GCS-based device is the nanocryotron (nTron), which is a three-terminal device where gate and S nanoconstriction are connected via a so-called choke. In an nTron, an applied $I_{bias}$ drives the device out of its superconducting state (inducing heating) and varies its resistance from zero to several MΩ [98]. Unlike what is expected for GCS devices, however, nTrons are slow to reset (since they are driven thermally) with $f_{max}$ ~ 1 GHz [103-104], they are hysteretic and have poor input-output isolation [103] – which are all drawbacks for logic applications.

Other superconducting devices recently proposed for logics include multi-terminal SFIFSIS (F being a ferromagnet and I an insulator) devices, where the $I_c$ of the JJ is controlled via injection of quasiparticles from the SFIFS part of the device [105]. Although these devices have high input-output isolation, they have the drawback of being sensitive to magnetic fields, unlike GCS devices.

Another category of $V_G$-driven superconducting devices is hybrid JJs, where $V_G$ is applied to a proximitized weak link made of non-S material. Possible weak links include semiconducting nanowires [106-107], graphene [108], or of a two-dimensional electron gases (2DEG) [109]. In principle, the materials used as weak links in these devices introduce additional steps in the fabrication process compared to GCS devices. GCS devices are in fact all made from the same S material (other than from a combination of Ss with weak links of other materials), which is also a refractory metal above $T_c$ and therefore easy to pattern/process.

Despite the more complex fabrication process, however, for devices based on gated Josephson junctions with Al (S) contacts and proximitized graphene as weak link – also known as Josephson field effected transistors (JoFETs) – Generalov and co-workers [110] have recently demonstrated large-scale reproducibility using wafer-scale CMOS-compatible processing. Similar high reproducibility has also been reported by Delfanazari et al. [111] who have fabricated 18 chips with a total of 144 gate-controlled Nb-2DEG-Nb Josephson junctions, where the 2DEG is based on a semiconducting $In_{0.75}Ga_{0.25}As/GaAs$ heterostructure.

These gated junctions based on material hybrids may also operate at $V_{G,offset}$ lower than the typical of $V_{G,offset}$ of GCS devices ($V_{G,offset}$ ~ 10 Volts in ref. [110] and ~ 0.56 V in ref. [111]), which makes them very appealing for applications.

## 5.2. Other applications of GCS devices for superconducting electronics

The GCS has also been explored recently [112] to obtain a gate-tunable superconducting diode effect (SDE). Similar to the GCS, the SDE has been the object of intensive studies over the past few years for both fundamental and technological reasons, and it consists in the observation of non-reciprocal transport in a superconducting device [113]. Specifically, the non-reciprocal transport corresponds to a different amplitude of $I_c$ in the superconducting device, depending on the polarity of the bias current $I_{bias}$ injected through the device, meaning that $|I_c^-| \neq I_c^+$ ($I_c^-$ and $I_c^+$ are the Ic extracted from the $I(V)$ curve of the device for negative and positive sweeping of $I_{bias}$, respectively). Assuming, for example, that $I_c^- > I_c^+$, the condition $I_c^- \neq I_c^+$ also implies that there exists an $I_{bias}$ range such that $|I_c^-| > I_{bias} > I_c^+$, within which the SDE device is in the resistive (superconducting) state when $I_{bias}$ has a negative (positive) polarity. Although a SDE device behaves as another simple logic element with two states (i.e., superconducting and resistive) defined by the $I_{bias}$ polarity, similar to conventional diodes based in semiconductors, a SDE device can also find other interesting applications for the realization of more complex devices like superconducting circulators or isolators [114-115].

The realization of a SDE normally requires a material system that breaks both inversion and time reversal symmetry, with the latter that can be achieved, for example, not only thanks to the presence of intrinsic sources (e.g., the presence of magnetic elements) but also via an applied magnetic field. For a full introduction to the SDE see ref. [113] and studies cited therein. Margineda and co-workers, however, have recently shown that the GCS can be used as a possible tool to induce a SDE in a gated superconducting constriction, without any needs for a source of time-reversal symmetry breaking. In their setup, $I_{bias}$ flows first through the gated S constriction and then in a resistor ($R_c$) connected in series to the device, as shown in Fig. 15a.

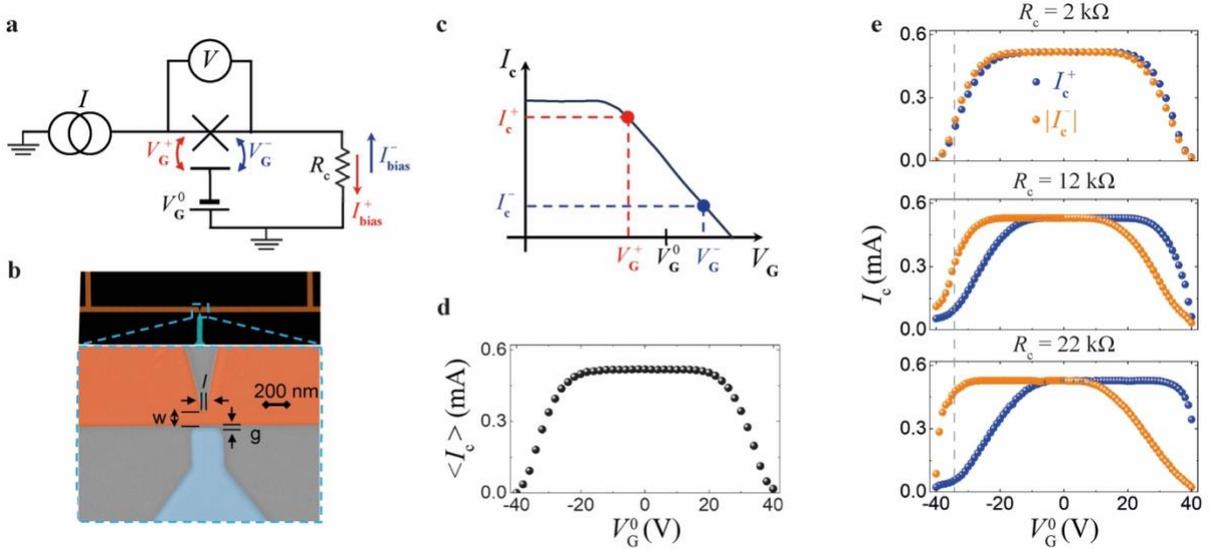

**Figure 15. Application of GCS for gate-controlled superconducting diodes.** (**a-b**) Schematic of a circuit used to implement a gate-controlled superconducting diode (a), and colored SEM image of an actual device used for its realization in (b). The schematic in (a) shows that the bias current $I_{bias}$ flows through a resistor $R_c$ and induces a shift in the applied voltage $V_G^0$ (measured in the absence of current), which depends on the bias current polarity (blue for positive and red for negative) such that $V_G^\pm = V_G^0 - I_{bias}^\pm R_c$. (**c-d**) The shift in the $V_G$ due to the voltage drop across $R_c$ results in a change in the working point of the GCS device on its $I_c(V_G)$ characteristic (c), with $I_c(V_G)$ shown in (d) for the average $I_c$ defined as $\langle I_c \rangle = (I_c^+ + |I_c^-|)/2$. (e) $I_c(V_G)$ curves for positive $I_c$, $I_c^+$, (blue curves) and amplitude of negative $I_c$, $|I_c^-|$, (orange curves) for different values of $R_c$ showing a large difference between $I_c^+$ and $|I_c^-|$ (as marked by dashed gray line). All panels are adapted from D. Margineda *et al*., ArXiv pre-print at https://arxiv.org/abs/2311.14503 (2023) [112]; licensed under a CC BY license.

The current flowing through $R_c$ induces a shift in the $V_G$ applied to the gate, which is different depending on the polarity of the same current, since for one polarity the voltage shifts adds up to the $V_G$ applied in the absence of current ($V_G^0$) whereas, for the opposite polarity, it subtracts from it (see red and blue labels in Figs. 14 a and c). The different voltages effectively applied as result of this shift ($V_G^+$ and $V_G^-$ in Fig. 15c) correspond to different working points on the $I_c(V_G)$ curve making $I_c^- \neq I_c^+$ – which is the condition required for a SDE to occur, as explained above. The large differences in $I_c^-$ and $I_c^+$ can be also visualized from Fig. 15e showing that the difference between the $I_c(V_G)$ curves measured for different $I_c$ polarities ($I_c^-(V_G)$ in orange and $I_c^+(V_G)$ in blue) increases as the value of $R_c$, and hence the voltage drop on this resistor, increases (see dashed grey line in the same Fig. 15e). In this paper, the authors have in fact reported values of the so-called rectification factor $\eta = \frac{|I_c^-| - I_c^+}{|I_c^-| + I_c^+}$ as large as 90%. $\eta$ is a parameter measured to quantify the performance of a diode because $\eta = 100\%$ corresponds to a superconducting diode that is fully resistive for a certain polarity of $I_{bias}$, and fully superconducting for the opposite polarity, provided that $I_{bias}$ does not exceed the superconducting critical current.

### 5.3. Applications of GCS devices for quantum computing

In addition to superconducting logics, GCS devices can also find application in emerging quantum computing technologies. GCS devices, for example, can be directly integrated as tunable elements into quantum processing units (QPUs) based on superconducting circuits – this type of QPU represents the leading platform for the realization of a universal gate-based quantum computer [116]. QPUs necessitate of tunable elements to reversibly switch on/off the interactions needed to control multi-qubit gates [117], to reset the qubits [118], and to decouple them from readout circuitry during their operation [119]. In state-of-the-art QPUs, such tunability is often achieved by employing flux-tunable elements such as SQUIDs or superconducting nonlinear asymmetric inductive elements [120]. However, flux control has problems with crosstalk (at the percent level) [121] and frequency-dependent transfer function of the control lines, requiring predistortion of baseband pulses [122]. The availability of a fast-tunable element based on local *E*-field control could largely mitigate such issues. This technological need has motivated the development of the hybrid JJ devices listed above [123]. It is still unclear, however, whether any of these hybrid JJs can be scaled to multi-qubit QPUs.

By contrast, prospective devices based on the GCS would be immediately scalable and compatible with state-of-the-art QPU fabrication recipes. It remains to be seen, however, whether sufficiently high switching speeds $f_{max}$ can be achieved (operation at the ns-level will be needed), and, perhaps most importantly, to what extent the integration of GCS elements affects the QPU coherence times [124].

The idea of developing superconducting qubits with tunable frequency has been recently proposed [125-126] and several approaches have been proposed for their realization. In the most common type of superconducting qubit known as transmon [127], which includes two S/I/S Josephson junctions on its branches (with I being an insulator), the qubit frequency depends on the ratio between the Josephson energy to the charging energy, which is in turn affected by the $I_c$ of the junctions. One approach recently proposed to have an additional knob to tune the frequency of transmon qubit consists in replacing one of the S/I/S junctions with an S/F/S/I/S junction, for which the $I_c$ can be modulated by switching the magnetization of the F layer via an applied magnetic field *B* [128]. Although the performance of such a device called ferrotransmons remains to be verified, the realization a ferrotransmon per se poses several fabrication-related challenges, in addition to the need for an applied *B* for the qubit control. If a gate-controlled junction or a GCS device were used instead as tunable element in a transmon, such an approach could be exploited to change the transmon frequency electrically (i.e., via an applied $V_G$) other than magnetically. It remains to be checked, however, whether and how the $I_{leak}$ induced by the applied $V_G$, which is usually present in GCS devices, negatively affect coherence of the qubit due to quasiparticle poisoning introduced in the circuit, as explained above.

Nonetheless, it is worth noting that, even if GCS elements were to perform worse than conventional JJs in transmon qubits, they could still be integrated in auxiliary modes of the QPU (e.g., tunable couplers, readout resonators, Purcell filters [129-130]), which can tolerate around 100 times higher losses than the modes hosting the computational qubits. Finally, even if direct integration of GCS elements into the QPU turned out to be problematic, GCS elements could still find application in other layers of the quantum hardware stack and be used, for example, to multiplex routing of microwave signals to the QPU [131-132], or as building blocks of quantum-limited amplifiers [133-135].

A possible integration of a GCS device into the resonator used for the qubit control/readout is shown in Fig. 16. In this case, the GCS device would be used as a gate-tunable element to change the frequency of the resonator (as done, for example, in ref. [18]) and bring it closer to the frequency of operation of the superconducting qubit or further away from it (in case decoupling of the resonator from the qubit is sought, for example, to keep the qubit state).

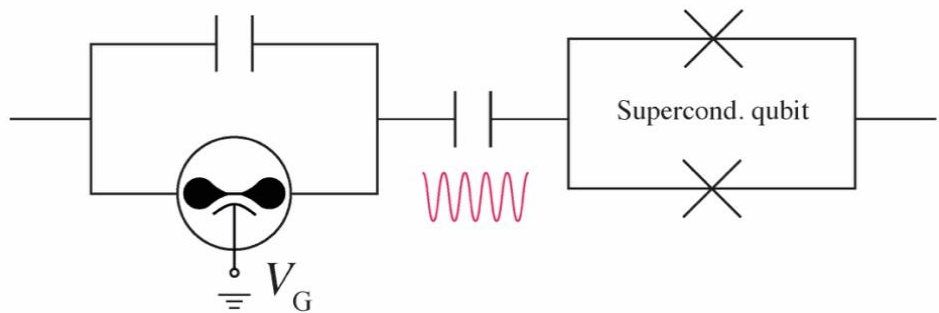

**Figure 16. Schematic of a resonator embedding a GCS device and coupled to a superconducting qubit.** The presence of the GCS device allows to vary the inductance (via the applied gate voltage $V_G$) and in turn shift the resonant frequency of the resonator, bringing it closer or further away from the qubit frequency.

Another potentially impactful application of GCS devices is related to their usage for efficient signal routing in QPUs, which can in turn enhance the control of qubits. GCS devices could be in fact integrated into flip-chip controllers, enabling precise signal routing in a cryogenic environment while maintaining qubit coherence [136]. For example, GCS radiofrequency switches based on superconducting materials can reduce the number of input/output cables, which traditionally limit scalability and introduce Joule heating. This approach minimizes power dissipation and signal loss, which are critical for maintaining low temperatures and improving the QPU uptime. Moreover, the use of GCS-based routing solutions allows multiple qubit configurations to be controlled via fewer RF lines, drastically improving the scalability of quantum systems [137].

## 6. Outlook and future challenges

Despite the large variation in S materials, device geometries and fabrication protocols across the studies done on the GCS, both intrinsic (e.g., SOC strength) and extrinsic properties of a S material (e.g., roughness, microstrain and surface) seem to influence the GCS. Confirming the existence of correlations between the GCS and some of these parameters is crucial not only to better understand the physics of the effect, but also to achieve fine control over it for future technological applications. A full understanding of the physics underlying the GCS remains the first objective to pursue in the future.

Although the studies carried out to date on the GCS show a large variation in the types of devices tested and experimental signatures found, most of these studies also share a common set of observations summarized in Table 1, which can be considered as common features of the GCS and include:

- Close to symmetric suppression of $I_c$ with $V_G$ polarity [12,14,16,17,20-22,26-29,31,35-36].
- Negligible or small variations of $V_{G,offset}$ with $T$ and applied $B$ [12,14,17,19,21-23,26,31-32,35-36].
- Non-null $I_{leak}$ (independently on its absolute value) present under an applied $V_G$ [12,14,16,17,20-24,26-29,31-36].

Any new theoretical models proposed to explain the GCS should also account for these features.

In addition to the above, there are other types of experimental features that have been considered as evidence in support of one or more scenarios amongst those proposed to date to explain the GCS (see also Table 3), which include:

- Enhancement in non-thermal phase fluctuations due to $V_G$ inferred from SCD measurements [21,31-32,50], considered as evidence for scenarios 3 or 4.
- Increase of quasiparticle population detected in tunneling devices [24,32,50], considered as evidence in support of scenarios 1 or 3.
- GCS also in devices with no substrate-mediated coupling between S and gate such as suspended nanowires or STM tip for tunneling current injection [22,52], considered as evidence for scenarios 4 or 1.

Amongst the mechanisms proposed to date to explain the GCS, some of them like field emission (scenario 1) and phonon heating (scenario 2) seem to be at play only in part of the experiments. At the same time, more than a single mechanism can be at play in some devices. For example, the $E$ and $I_{leak}$ generated by the applied $V_G$ may both affect the phase of the S, which also suggests that their contributions can be difficult to disentangle in some devices. In devices where $I_{leak}$ has been minimized using for example ILG, a partial and not full $I_c$ suppression has been observed [25].

Most of the proposed mechanisms are also related in some form to $I_{leak}$, which suggests the need for a standardization in its measurement to properly compare the behavior of different devices. $I_{leak}$ is always present in any devices made on a dielectric substrate with a side gate or with a top gate, independently on how small $I_{leak}$ can be. This is because an applied $V_G$ always builds up accumulation of charges at the S/substrate interface, where the presence of pinholes or defects can create percolating paths and make the accumulated charges flow in S. Also, in several experiments on the GCS, $V_{G,offset}$ is often close to the breakdown voltage of the dielectric substrate and/or can induce electromigration, which can induce changes to the dielectric including SILC effects as reported in ref. [16,29] that in turn affect $I_{leak}$. This is why a reduction in $V_{G,offset}$ would be beneficial also to reduce substrate contributions to $I_{leak}$.

Together with a better understanding of the mechanism responsible for the GCS, a reduction in $V_{G,offset}$ remains the second major challenge to tackle in the research field of the GCS because a lower $V_{G,offset}$ would not only reduce $I_{leak}$, and possibly improve device performance, but also make interconnection of GCS devices as well as their interfacing with CMOS easier.

Recent experiments [34], however, also suggest that a large $I_{leak}$ (> 10 nA) alone is not sufficient for a GCS, but other physical parameters, possibly related to the S surface states and varying depending on the surface is treated [138-139], also concur toward the GCS observations. To establish the importance of surface states, further studies should be carried out, where parameters like disorder and surface roughness, which can affect surface states, are systematically varied by changing the growth conditions of the S. Preliminary results showing the importance of these parameters have been reported in ref. [36] as discussed above, but more systematic studies are needed. The importance of surface states can also be assessed by using Ss with more complex band structure or high SOC or unconventional magnetic surface states (e.g., A15-type S like $Nb_3Ge$ or metal-oxide S like $Sr_2RuO_4$) [140-143]. If surface states play a crucial role, a GCS should be observed also in devices made with a top-down approach (as done in ref. [36] for NbRe devices) based on these Ss.

Spectroscopy studies are essential at this stage to draw the complete picture on the GCS. Spectroscopy techniques like nano angle-resolved photoemission spectroscopy can be used to study the evolution of surface states under an applied $V_G$, both in devices with and without a GCS, and to understand their importance. Other techniques like SQUID-on-tip [144] or nitrogen vacancy (NV) magnetometry [145] should be used to study how the spatial distribution of currents (and associated magnetic fields) inside a S nanoconstriction changes under a $V_G$. This can help understand whether parts of the constrictions (e.g., those closer to the gate electrode) turn into the normal state at lower $|V_G|$ than others, or if the supercurrent becomes weaker, as $|V_G|$ is progressively increased, near defects or other parts of the constriction like sharp corners where current crowding effects can be more relevant. Muon spectroscopy experiments can also help study how the screening current distribution in a S varies [146-148] under an applied $V_G$.

More low-$T$ STM studies, where local DoS spectra are acquired by STM both with and without $V_G$, can be also helpful onto the physics of the GCS. This type of STM studies that are still lacking at the moment would require a customized STM setup, where the gated device can be not only located and brought within the scan area of the piezoelectric tube (typically ~ 1 μm x 1 μm), but which should also be equipped with lines to apply an $I_{bias}$ and/or a $V_G$. Scanning gate microscopy (SGM) can also give important insights [149-150]. By applying a strong $E$ to the S nanoconstriction with a charged atomic force microscope tip ($E$ is strong due to the small tip-to-sample distance), SGM would allow to investigate, with sub-nanometer resolution, how a strong $E$ affects the transport properties of a device.

Measurements of the dynamics of gate-controlled superconducting devices with the determination of their highest switching frequency $f_{max}$ represents the third major goal to pursue, which is crucial also to understand the full technological potential of GCS devices. In an earlier report [23], it was shown that the GCS can follow a dynamic $V_G$ excitation with a frequency of ~ 10 MHz, although the actual $f_{max}$ of the devices was not quantified. More recently, Joint and co-workers have performed a first detailed characterization of the dynamic response of Nb Dayem bridges embedded in λ/4 superconducting microwave resonators [30]. In their study, they have found that the switching response of the devices is strongly dependent on the type of gate electrode and can reach switching frequencies above 500 MHz for devices with remote electrodes (i.e., placed ad a $d_{gate}$ ~1 μm) compared to devices with closer finger-type electrodes for which the switching frequencies are much lower (~ 60 MHz). The different behavior of the two types of devices is ascribed to different mechanisms activated by $V_G$, with injections of quasiparticles from the gate into the S constriction that is more significant for the configuration with closer electrodes, compared to devices with remote gates where a flux of phonons induced by $V_G$ is considered as the main mechanism affecting the switching dynamics.

Although the same authors of ref. [30] state that the $f_{max}$ of their GCS devices with remote gates could be higher than 500 MHz and may require lines with broader bandwidths and on-chip filtering to be fully resolved, this study [30] shows that future measurements of the switching dynamics of GCS devices are important not only for technological applications, but also because they may shed light onto the physics of the GCS. If phonon-induced heating were dominant for the GCS, for example, then $f_{max}$ would be probably limited to a few GHz by thermal effects and by quasiparticle recombination times (< 100 ps) [23,97]. In devices where phonon-heating is not at play, but mechanisms like phase fluctuations or field effect are more relevant, $f_{max}$ could be much higher and reach hundreds of GHz depending on the S material since both mechanisms could act on the phase of the S condensate (albeit in different ways). More measurements of the switching dynamics of GCS devices on the model of those reported in ref. [30] remain therefore a crucial objective to pursue soon.

Another important objective to pursue soon, as discussed above, consists in the increase in $V_{out}$. Although preliminary efforts in this direction have been made in ref. [29] where $V_{out}$ ~ 0.25 V have been reached in wide Nb bridges, larger values seem totally within reach if Ss with higher critical current density and/or normal-state resistivity were adopted. Similarly, different device geometry (e.g., with longer S constrictions) would also help reach this goal. The increase in $V_{\text{out}}$ should occur simultaneously with the reduction in $V_{\text{G,offset}}$, because these voltages should be of the same order of magnitude to allow interconnection of GCS devices and an increase in the fan-out.

## 7. Conclusions

To summarize, systematic studies on the effect of individual material parameters in combination with spectroscopy may prove crucial in the next years to understand the exact physical origin of the GCS. This fundamental understanding should proceed alongside with the identification of the best S materials, device geometries and fabrication protocols to ensure a high reproducibility of the effect, possibly at much lower $V_{\text{G,offset}}$ than the typical values reported to date. The dynamic switching of GCS devices up to $f_{\text{max}}$ of hundreds of GHz, the device control with top gates, and the development of basic circuits using protocols based on subtractive patterning (for higher device scalability) are all important milestones to reach to develop technologies based on the GCS. If all these milestones are achieved, GCS-based devices can have a potentially disruptive impact on future technologies for superconducting electronics and quantum computing with superconducting qubits.

## Acknowledgements

This work was funded by the EU's Horizon 2020 research and innovation programme under Grant Agreement No. 964398 (SUPERGATE).

## Competing interests

The authors declare no competing interests.